\documentclass{article}
\usepackage[square,numbers]{natbib}
\usepackage{amsmath}
\usepackage{hyperref}
\PassOptionsToPackage{numbers, compress}{natbib}
\usepackage{comment}
\usepackage[preprint]{neurips_data_2024}

\usepackage[utf8]{inputenc} 
\usepackage[T1]{fontenc}    
\usepackage{url}            
\usepackage{booktabs}       
\usepackage{amsfonts}       
\usepackage{nicefrac}       
\usepackage{microtype}      
\usepackage{xcolor}         
\usepackage{graphicx}       

\definecolor{citecol}{HTML}{6F130C}
\definecolor{tableofcontent}{HTML}{1F4A83}
\definecolor{urlcol}{HTML}{2470D8}

\usepackage{hyperref}
\hypersetup{
    colorlinks=true,       
    linkcolor=tableofcontent,
    citecolor=citecol,        
    urlcolor=blue,           
}

\title{
A Fine-tuning Dataset and Benchmark for Large Language Models for Protein Understanding
}

\author{%
  Yiqing Shen\textsuperscript{1,2,$\dagger$}, Zan Chen\textsuperscript{1,$\dagger$},  Michail Mamalakis\textsuperscript{3,$\dagger$}, Luhan He\textsuperscript{1,$\dagger$}, Haiyang Xia\textsuperscript{1,4,$\dagger$}, 
  \\\\
  {\bf Tianbin Li\textsuperscript{5}, Yanzhou Su\textsuperscript{5}, Junjun He\textsuperscript{5}, Yu Guang Wang\textsuperscript{1,5,6,7}}\thanks{Corresponding author, E-mail: yuguang.wang@sjtu.edu.cn. \textsuperscript{$\dagger$}Equal contribution.}\\\\
  \textsuperscript{1}Toursun Synbio \qquad
  \textsuperscript{2}Johns Hopkins University \qquad
  \textsuperscript{3}University of Cambridge\\\\
  \textsuperscript{4}Shanghai Institute for Biomedical and Pharmaceutical Technologies\\\\
 \textsuperscript{5}Shanghai AI Laboratory \qquad
  \textsuperscript{6}Shanghai Jiao Tong University
  \qquad
  \textsuperscript{7}UNSW Syndey\\
}

\definecolor{newcolor}{rgb}{.8,.349,.1}

\begin{document}

\maketitle

\begin{abstract} 
The high similarities between protein sequences and natural language, particularly in their sequential data structures, have driven parallel advancements in deep learning models for both domains. 
In natural language processing (NLP), large language models (LLMs) have achieved remarkable success in tasks such as text generation, translation, and conversational agents, owing to their extensive training on diverse datasets that enable them to capture complex language patterns and generate human-like text. 
Inspired by these advancements, researchers have attempted to adapt LLMs for protein understanding by integrating a protein sequence encoder with a pre-trained LLM, following designs like LLaVa.
However, this adaptation raises a fundamental question: \textit{``Can LLMs, originally designed for NLP, effectively comprehend protein sequences as a form of language?''}
Current datasets fall short in addressing this question due to the lack of a direct correlation between protein sequences and corresponding text descriptions, limiting the ability to train and evaluate LLMs for protein understanding effectively.
To bridge this gap, we introduce \texttt{ProteinLMDataset}, a dataset specifically designed for further self-supervised pretraining and supervised fine-tuning (SFT) of LLMs to enhance their capability for protein sequence comprehension.
Specifically, \texttt{ProteinLMDataset} includes 17.46 billion tokens for pretraining and 893K instructions for SFT. 
Additionally, we present \texttt{ProteinLMBench}, the first benchmark dataset consisting of 944 manually verified multiple-choice questions for assessing the protein understanding capabilities of LLMs.
\texttt{ProteinLMBench} incorporates protein-related details and sequences in multiple languages, establishing a new standard for evaluating LLMs' abilities in protein comprehension.
The large language model InternLM2-7B, pretrained and fine-tuned on the \texttt{ProteinLMDataset}, outperforms GPT-4 on \texttt{ProteinLMBench}, achieving the highest accuracy score.
The dataset and the benchmark are available at \url{https://huggingface.co/datasets/tsynbio/ProteinLMDataset/} and
\url{https://huggingface.co/datasets/tsynbio/ProteinLMBench}.
The code is available at \url{https://github.com/tsynbio/ProteinLMDataset/}.
\end{abstract}

\section{Introduction} 
Protein science offers insights into biological processes at the molecular level, driving progress in medicine, biotechnology, and our understanding of life. 
Given the similar sequential data structures of protein sequences and natural language, there have been parallel advancements in deep learning models, such as protein language models \cite{rives2019biological} and large language models (LLMs) \cite{gpt4}.
LLMs have already demonstrated their strong capabilities in text understanding across various tasks, prompting researchers to explore their potential in protein understanding from the perspective of multi-modal language models (MMLMs) by integrating protein sequences or structures with textual content using separate encoders.
%
%
Specifically, they encode each modality (protein sequence or protein structure) independently before combining them with a frozen LLM \cite{xu_protst_2023,zhou_protein_2023}. 
However, this approach poses challenges in fully leveraging the intricate connections between protein sequences and textual information.
As far as we can ascertain, there is no comprehensive dataset that seamlessly integrates protein sequences and their corresponding textual descriptions, enabling effective training of language models to comprehend protein information.
This gap limits the ability to train LLMs to fully utilize the intricate connections between protein sequences and textual information.

To address these limitations, we introduce a comprehensive large-scale protein sequence and text (seq-text) hybrid dataset named \texttt{ProteinLMDataset}. 
This dataset is designed to train LLMs to learn the correspondence between textual descriptions and protein sequences, enabling them to effectively understand and leverage the intricate connections between these two forms of data. 
Furthermore, we propose a novel benchmark \texttt{ProteinLMBench} comprising meticulously curated multiple-choice questions to rigorously evaluate the LLMs' proficiency to comprehend protein sequences.
To the best of our knowledge, this is the first dataset of its kind, combining the largest dataset for both self-supervised learning (SSL) and supervised fine-tuning (SFT) specifically in the protein science domain, along with the first manually annotated multiple-choice questions benchmark for evaluating protein understanding capability with LLMs.

The \texttt{ProteinLMDataset} comprises 17.46 billion tokens for self-supervised learning and 893K instructions for supervised fine-tuning.
\texttt{ProteinLMBench} contains 944 manually annotated multiple-choice questions for evaluation.
The self-supervised dataset is structured into three segments: 0.69\% Chinese-English text pairs in protein science, 41.51\% protein sequence-English text pairs, and 57.80\% protein-related English text.
The supervised fine-tuning dataset includes seven tasks: a novel enzyme chain of thought (ECoT), protein functionality, induction, disease involvement, post-translational modifications, sub-unit structure, and tissue specificity. 
The ECoT approach aims to innovatively enable LLMs to think step-by-step and generate well-founded protein knowledge, thereby enhancing the reliability and accuracy of outputs in protein science and engineering \cite{wei2022chain}.
Unlike the plain chain of thought, ECoT incorporates prior knowledge of protein science by guiding LLMs to infer protein functions based on the reactions they are involved in. This approach enhances logical reasoning capabilities in the protein domain.
\texttt{ProteinLMDataset} aims to bridge existing dataset gaps and unlock the potential of LLMs in protein data analysis by seamlessly integrating protein sequences with their relevant textual information.

The major contributions of this work are three-fold.
Firstly, we present the first large-scale protein-text dataset designed to enable LLMs to comprehend protein sequences without an extra encoder.
Secondly, we introduce the Enzyme Chain of Thought (ECoT), a novel mechanism specifically for protein understanding, allowing LLMs to generate reliable and accurate protein knowledge step-by-step.
Thirdly, we introduce \texttt{ProteinLMBench}, the first comprehensive, manually annotated benchmark for thoroughly evaluating LLMs' comprehension and representation of protein sequences.

\section{Related Work} \label{related_work}
\label{rw}
An extended version of the literature review and related work is presented in supplementary material: 'Thoroughly Related Work'. In this section, we summarize the main limitations found in the literature.
\par \textbf{Existing Chinese-English datasets.} Traditional benchmarks in the English language primarily focus on evaluating specific LLM abilities related to singular tasks, such as providing answers in general question answering, text summarization, or general linguistic understanding \cite{1,2,3}. The existing literature on English benchmarks primarily focuses on specific model abilities for singular linguistic tasks, with limited extension to multi-domain and multi-task assessments \cite{4,5,6}. However, in the context of Chinese NLP benchmarks, significant limitations emerge, including rational biases with geographic restrictions to China regions dataset collection, absence of analytical documentation and statistics, and lack of interpretability with limited insights into reasoning abilities \cite{7,8}. Overall, these limitations underscore the need for more comprehensive and diverse multi-task language datasets, addressing a broader geographic scope of collection, incorporating varied evaluation metrics, and ensuring coverage of distinct writing styles, including scientific content. Notably, there is currently no dataset known to the authors that provides text-to-text pairs for English and Chinese translation.
\par \textbf{Existing protein sequence datasets.} Several crucial datasets related to protein sequences, including UniProtKB \cite{14}, PEER \cite{19}, RefSeq \cite{15}, PDB \cite{11}, CATH (Class, Architecture, Topology, Homology) \cite{10}, and CASP15 \cite{12,13}, cover a broad spectrum of different protein sequences. However, despite their breadth, limitations exist, such as the focus on sequence-based data in the PEER dataset \cite{19}. Extending beyond these datasets presents challenges, highlighting the need for ongoing community collaboration to ensure sustained effectiveness. The literature reveals a lack of representation of the full diversity of protein structures, as evidenced by limitations in RefSeq \cite{11}, and limited applicability for real-world tasks, as noted in \cite{icrl24}. Addressing these limitations requires further research to optimize instruction tuning methods, refine the model's ability to understand and follow instructions, and improve its generalization across various protein and biomolecular tasks, as highlighted in previous works \cite{11,12,13}. Besides, it is essential to acknowledge potential biases inherent in community contributions, which may affect the representativeness of the dataset. The utilization of automated systems introduces the possibility of errors, impacting the accuracy of annotations, as noted in the limitations of UniProtKB \cite{14}. Furthermore, annotating proteins that are yet to be characterized presents inherent challenges, with limitations persisting in achieving comprehensive characterization, as observed in the PEER dataset \cite{19}. Managing the influx of whole-genome sequencing data and integrating diverse external resources also present significant challenges, potentially complicating the coherent presentation of data.
\par \textbf{Protein design databases.} KEGG \cite{16} serves as a curated resource, offering researchers a comprehensive repository of accurately annotated biological entities, including molecular networks, genes, proteins, chemicals, and health-related data. Despite its strengths, KEGG has limitations, such as a predominant focus on disease-related perturbations in human cells, potential biases in data inclusion, and the resource-intensive nature of manual curation. In contrast, the STRING database \cite{17} excels in its extensive integration of known and predicted protein associations, covering both physical interactions and functional relationships across over 14,000 organisms. However, STRING faces challenges like potential biases in text mining, reliance on predictive methods, and the lack of options for tissue-specific network pruning. Lastly, InterPro \cite{18}, a leading protein classification database, provides comprehensive annotations through the integration of predictive models, offering insights into families, domains, and conserved sites. Despite its valuable contributions, InterPro's limitations include infrequent updates, challenges in interpretability, and concerns about sustainability.

\section{Dataset and Benchmark Composition}
\subsection{ProteinLMDataset}
The \texttt{ProteinLMDataset} includes both self-supervised and supervised fine-tuning components.
The self-supervised part is strategically categorized into three primary segments, each contributing distinct dimensions to our objectives of empowering the LLMs to understand protein sequence.
The first segment focuses on comparing biological science in Chinese and English, encompassing 7,000 entries and yielding over 120 million tokens for training and tuning purposes.
%
The second segment contains protein sequence and English text pairs, derived from a weakly linked PMC full-text dataset. Associations are established using the Structure Integration with Function, Taxonomy, and Sequence (SIFTS) database (\url{https://www.ebi.ac.uk/pdbe/docs/sifts/}), which maps PDB to UniProt IDs and subsequently to protein names, guiding our search for relevant literature within the PMC database. 
%
%
The full-text component comprises 320,000 entries, providing an extensive corpus exceeding 2.9 billion tokens.
Moreover, the PubMed abstract contributes over 30 million entries and 7.0 billion tokens for comprehensive training and tuning. 
For protein sequence and text pairs, we leverage a highly relevant PMC full-text dataset totaling 18,000 entries and surpassing 195 million tokens for nuanced training and tuning. 
The integration of the UniProtKB Swiss-Prot dataset further enriches this category with an additional 349 million tokens.
The final segment involves extracting insights from a PubMed abstract dataset, boasting more than 3 million entries, and generating over 7.0 billion tokens for robust training and tuning. 
This strategic categorization ensures a diverse and versatile dataset, addressing the limitations of existing benchmarks and facilitating advancements in cross-lingual language models and protein sequence understanding.

The supervised fine-tuning component in \texttt{ProteinLMDataset} is a collection of 893,000 instructions, spanning seven different segments: enzyme Chain of Thought (10.8k), protein functionality (465k), induction (25.4k), disease involvement (5.58k), post-translational modifications (45.8k), sub-unit structure (291k), and tissue specificity (50.3k). 
It includes analytical instructions, input tokens, and ground truth prediction outputs for fine-tuning and performance score determination. 
The primary source of data collection is UniProtKB.

\subsection{ProteinLMBench}
Our proposed benchmark, \texttt{ProteinLMBench}, features an evaluation dataset consisting of 944 six-choice questions, each accompanied by an explanation of the correct answer.
These questions cover a range of topics including protein-based property prediction, protein descriptions, and protein sequence understanding.
This benchmark is designed to assess the ability of large language models to interpret and analyze protein sequences in conjunction with their associated textual descriptions.

\section{Collection Methodology}
\label{m}

\subsection{The self-supervised dataset collection}
\paragraph{Biology Chinese and English text pair collection.} 
We extracted Chinese-English text pairs from articles published in the \textit{Chinese Journal of Biotechnology} and the \textit{Chinese Journal of Cell Biology}. 
This extraction process ensures a diverse and authentic representation of scientific language. 
The selected time interval, 2013-2024, enhances the dataset's relevance by capturing contemporary scientific discourse in both Chinese and English languages. 

\begin{figure}[ht]
    \centering
    \includegraphics[width=0.4\textwidth]{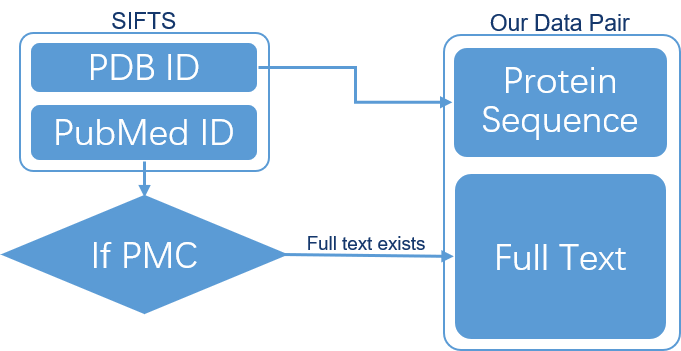}
    \caption{Process for pairing proteins with the corresponding PMC articles' text}
    \label{fig:1}
\end{figure}

\paragraph{Protein sequence and English text pair.}
We retrieved Protein Data Bank (PDB) IDs of proteins and their corresponding PubMed IDs from the SIFTS database (\url{https://www.ebi.ac.uk/pdbe/docs/sifts/}). Using a weakly linked PMC full-text dataset, we established associations through the \textit{protein\_name\_to\_pmcid.json} index file derived from \textit{pdb\_chain\_uniprot.csv}. This mapping, connecting PDB to UniProt IDs and subsequently to protein names, guided our literature search within the PMC database. In this manner, we can associate a specific protein sequence with the complete texts of the most relevant articles in PMC.
We examined the PubMed articles for PubMed Central (PMC) IDs; when available, we accessed the full text. This enabled us to acquire protein sequences using PDB IDs and retrieved detailed research articles on the proteins, creating pairs to describe the protein sequences in human language (Fig. \ref{fig:1}).

\begin{figure}[ht]
    \centering
    \includegraphics[width=1.\textwidth]{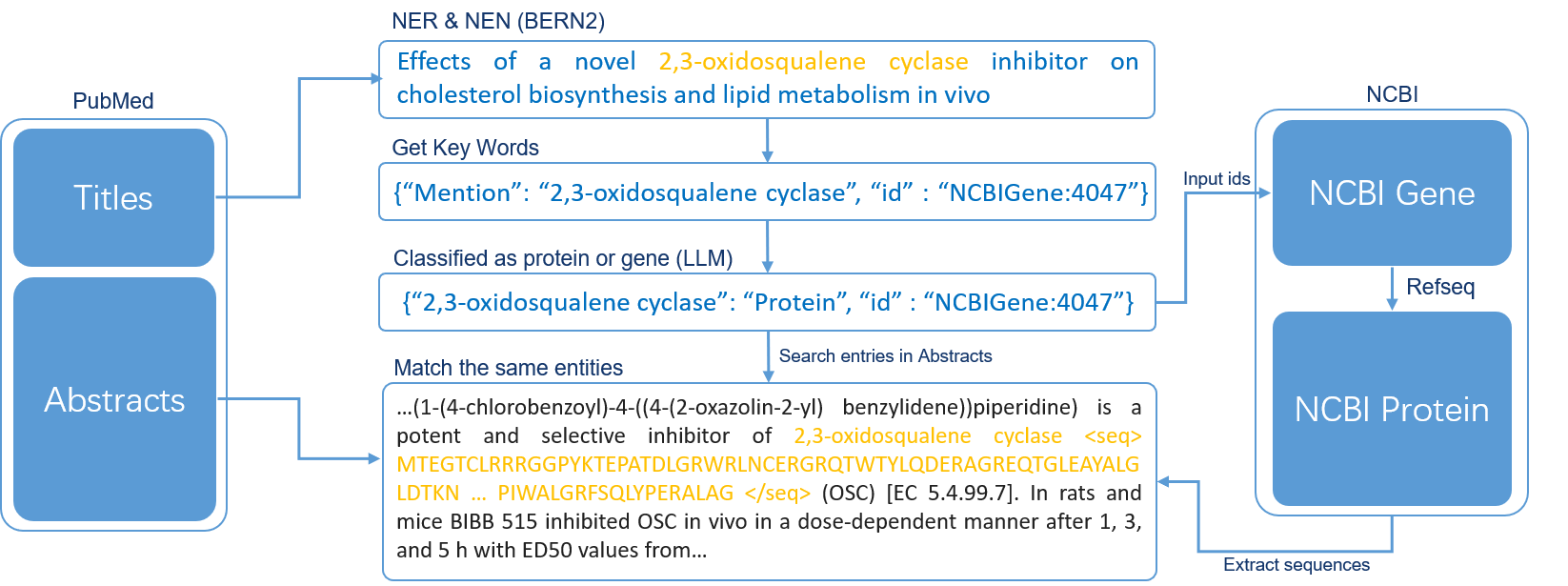}
    \caption{Process for Inserting Protein Sequences into Text}
    \label{2}
\end{figure}
\par \textbf{Protein sequence collection.} We employed BERN2 \cite{bern} which is a Named Entity Recognition (NER) model to extract NERs from the titles and abstracts of all PubMed articles, identifying nouns potentially referring to proteins or genes (since it is challenging to distinguish whether a single word refers to a protein or gene without context) and recorded their confidence scores for further analysis (Fig. \ref{2}).
We then linked all extracted nouns to specific entities in the NCBI Gene database \cite{ncbi} through Named Entity Normalization (NEN) and obtained their NCBI Gene IDs. For these Gene IDs, we could map them to specific protein sequences via the NCBI's RefSeq database.
Afterward, we extracted all article titles and the results from the NER \& NEN models with a confidence score above a certain threshold (here, 98\%), inputting them into a large language model to determine whether these nouns are referred to as proteins or genes in the given context. We assumed that protein names appearing in titles are likely the focus of the research and used the abstracts to help describe these proteins.
Finally, we located the first occurrence of each noun in the abstracts that matched the protein names identified in the titles with the same NCBI Gene ID and inserted the previously obtained protein sequences after these nouns. This way, we created a dataset of textual descriptions wrapping protein sequences.

\subsection{The supervised fine-tuning dataset collection}
The primary source for the supervised fine-tuning dataset is UniProtKB. Through a comprehensive analysis of the locally downloaded raw data, we identified a substantial corpus of curated and verified protein information. This encompasses a diverse range of attributes, including protein functionality, induction, disease associations, post-translational modifications, sub-unit structures, and tissue specificity.
For each protein, we crafted a dedicated template. The protein sequence was embedded within the question template, while its corresponding properties were incorporated into the answer template. This approach facilitated the creation of effective question-answer pairs to understand protein and its properties. In total, we generated approximately 883k such data entries.
\par To create the Chain of Thought (CoT) data, we utilized the IUBMB Enzyme Nomenclature as the raw source. We extracted the name, nickname, reaction formula, and comments (typically describing the enzyme's function) for each enzyme entry. The response generation process followed a systematic thinking process: first, the model outputted the reactions the enzyme would catalyze, and then provided the enzyme's comments, forming a chain of thought. During data creation, we generated multiple suitable templates and inserted the obtained information, resulting in an effective CoT supervised fine-tuning dataset comprising 10,800 entries.

\subsection{The benchmark multi-choice questionnaire collection}
The evaluation questionnaire, \texttt{ProteinLMBench}, comprises six choices for each question, accompanied by an explanation of the correct answer. The questions span various domains, including protein-based property prediction, protein descriptions, and protein sequence understanding. The questionnaire consists of a synthetic dataset generated by Retrieval-Augmented Generation (RAG; \cite{rag}) and GPT-4 \cite{gpt4}. We implemented a two-round verification scheme involving RAG to generate the initial questions, choices, and answers, followed by GPT-4 to validate the answers. Questions with inconsistencies between the two rounds were discarded, ensuring the consistency, robustness, and reproducibility of the questionnaire. Despite starting with 1,000 questions, the verification process yielded a final set of 944 questions.

\subsection{Text input filtering and tokenization procedure}
To preprocess the text inputs from three collections, we employed a filter cleaner to eliminate citations, references, tables, figures, commas, and any symbols unrelated to the task of interest. This filter was systematically applied to both titles and abstracts within the collected manuscripts. The resulting filtered outputs were stored in \textit{".json"} format. Subsequently, this \textit{json} files serve as input for extracting tokens of various sizes associated with the collection and the task of interest, such as protein sequences, pairs of English and Chinese text, protein descriptions, and design elements. The implementation source code for the filters is available on our GitHub repository at the following link: \url{https://github.com/tsynbio/Protein_LM/blob/main/src/PMC_data_collector.py}.
The tokenization is performed using the open-source coding of the \textit{Interlm} tokenizer based on byte-level byte-pair-encoding (\url{https://huggingface.co/internlm/internlm-7b/blob/main/tokenization_internlm.py}), which processes the filtered text inputs.

\section{Quality of Dataset and Benchmark}
\begin{figure}[t]
\centering
\centerline{
    \relax \textbf{a.}
    \includegraphics[trim={0.0cm 0.5cm 0.0cm 0.0cm},clip,scale=.275]{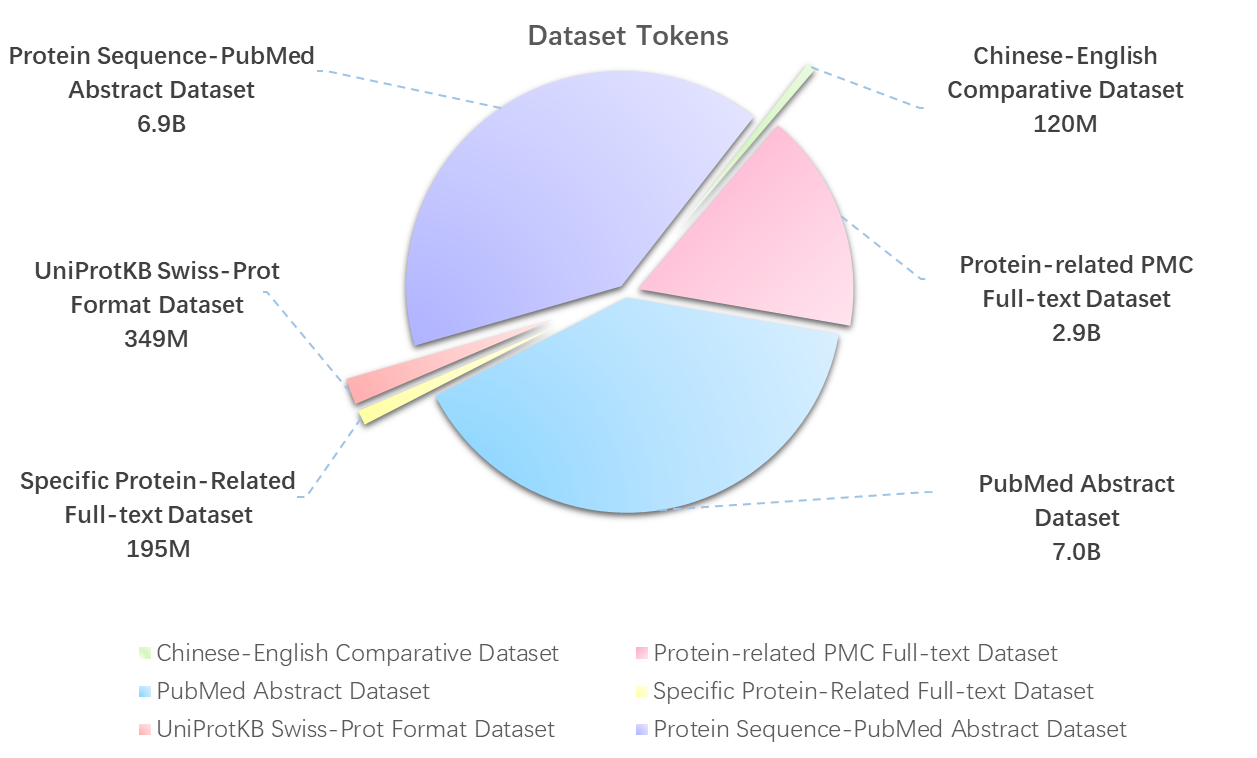}}
    \centerline{
    \relax \textbf{b.}
        \includegraphics[trim={0.0cm 8.5cm 10.0cm 0.0cm},clip,scale=.5]{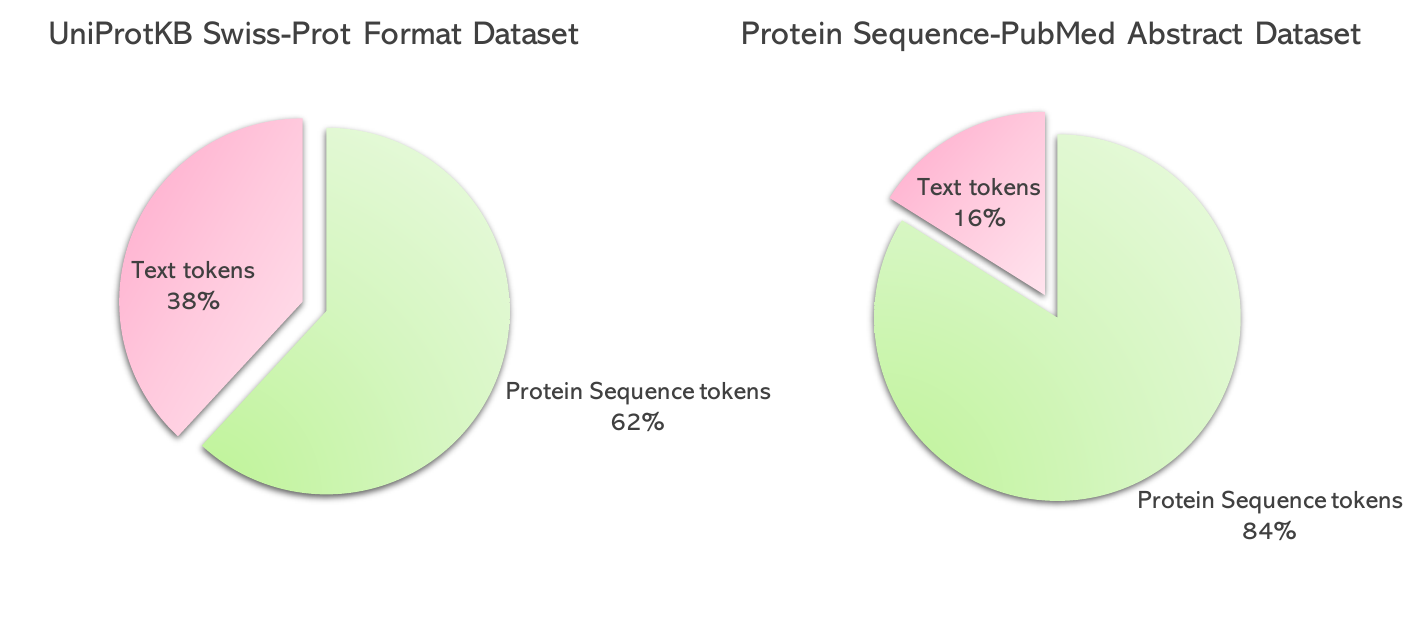}}
    \caption{Statistics of the sources and ratios of the three segments of \texttt{ProteinLMDataset}: Chinese-English pairs, protein sequences, and protein sequence-text pairs.}
    \label{fig:2}
\end{figure}

\paragraph{Diversity and statistics of ProteinLMDataset.}
The diversity and multi-source curation of the dataset are presented in Tab. \ref{t2}, highlighting the document collection based on the token length. The three segments of the proposed dataset includes manuscripts, abstracts, protein sequences, and taxonomy, divided into five collections: 'Biology Chinese/English text pair', 'PMC full-text manuscripts', 'PubMed Abstracts', 'UniProtKB Swiss-Prot', and 'Protein Sequence/PubMed Abstracts' and a variability ranges from minimum lengths of 49 to 549 characters to maximum lengths of 39,239 to 8,390,807 characters. Tab. \ref{t2} highlights the extracted tokens from the documentation characters after the filtering and noise reduction process (citations, references, tables, etc. unrelated to the task of interest). The produced tokens exhibit variability, ranging from minimum lengths of 21 to 101 characters to maximum lengths of 1,017 to 2,351,194 characters. These variations across different tokens, texts and combinations validate the unbiased and comprehensive content of our dataset.
\begin{table}[t]
  \caption{The overall statistics of extracted tokens from the proposed self-supervised dataset. The numbers are representation of the Length (Len.) token per document. The dataset classify in three segments-Biology Chinese/English (Chi/Eng) text pair, Protein Sequence and English (Seq/Eng) text pair and Protein Sequence (Seq)-.  The statistics are aggregated over the entire cohort.}
  \label{t2}
  \centering
    \begin{tabular}{lllll}
    \toprule
    \multicolumn{5}{c}{Tokens statistic of the self-supervised dataset}                   \\
    \cmidrule(r){1-5}
    Segment& Collection& Min Len. & Average Len. & Max Len. \\
    \midrule
    Biology Chi/Eng text pair&  Chinese/English text &  54/48  & 245/267   &  1017/1165 \\
    Protein Seq/Eng text pair & PMC Full text     & 101 & 9954 &  2351194 \\
    & PubMed Abstract     & 21 & 282 &  13983 \\
    & UniProtKB Swiss-Prot  & 69 & 591 & 35751  \\
    {Protein Seq} & Protein Seq      & 65 & 1650 & 70500 \\
    \bottomrule
  \end{tabular}
\end{table}
Fig. \ref{fig:2} illustrates the statistics regarding the different sources and ratios within our three segment datasets. Our dataset predominantly represents protein sequences, with the primary source being PubMed abstracts (6.9B), followed by UniProtKB Swiss-Prot collection (349M). Regarding protein design and information, the primary resources are derived from PubMed abstracts of scientific manuscripts (7.0B), followed by weakly linked PMC full manuscripts and highly relevant PMC full texts (2.9B; see Fig. \ref{fig:2}a.). Additionally, the Chinese-English collection comprises a substantial of 200M tokens. To delve deeper, the UniProtKB Swiss-Prot collection consists of 62\% protein sequences and 38\% text relevant to protein design and scientific biology documentation tokens, contrasting with the Protein Seq/PubMed Abstract collection, which includes 16\% text documentation and 84\% protein sequences tokens (see Fig. \ref{fig:2}b). 
\par Tab.~\ref{t31} presents the diversity of multi-design topics and curation of the fine-tuning dataset, highlighting the different tasks and analyzing the statistics based on token length. The proposed benchmark comprises seven segments: 'Enzyme CoT', 'UniProt Function', 'UniProt Induction', 'UniProt Involvement in disease', 'UniProt Post-translational modification', 'UniProt Sub-unit structure', and 'UniProt Tissue specificity'. These segments exhibit a wide range of token lengths, varying from a minimum of 65 to 88 characters to a maximum of 2,310 to 70,500 tokens. Tab.~\ref{t31} highlights the extracted token counts for each segment. After filtering and noise reduction, the instructions and output of the benchmark are provided on the website \url{https://huggingface.co/datasets/tsynbio/ProteinLMBench}. The variations across different token lengths, fine-tuning tasks, and combinations validate the unbiased and comprehensive nature of our benchmark.
\begin{table}[ht]
  \caption{The overall statistics of extracted tokens from the proposed fine-tuning dataset represent the token length per document. The dataset is categorized into seven segments: Enzyme Chain of Thought (ECoT), Protein Functionality, Induction of Protein Expression, Disease Involvement, Post-Translational Modifications, Sub-Unit Structure, and Tissue Specificity. These statistics are aggregated across the entire cohort.}
  \label{t31}
  \centering
  \resizebox{\textwidth}{18mm}{
    \begin{tabular}{lllll}
    \toprule
    \multicolumn{5}{c}{Tokens statistic of the fine-tuning dataset}                   \\
    \cmidrule(r){1-5}
    Segment & Scope& Min Len. & Average Len. & Max Len. \\
    \midrule
    Enzyme CoT &  molecule's expression  & 84 & 800 & 15300 \\
    UniProt Function & protein functionality  & 74 & 3610 & 70500 \\
    UniProt Induction & protein induction  & 82 & 1650 & 31300  \\
    UniProt Involvement in disease & protein diseases  & 88 & 3495 & 68800  \\
    Uniprot Post-translational modification & protein translation  & 88 & 361 & 6420 \\    
    UniProt Sub-unit structure & protein structure  & 74 & 3560 & 70500\\
    UniProt Tissue specificity & tissue gene expression  & 65 & 145 & 2310 \\
    \bottomrule
  \end{tabular}}
\end{table}

\paragraph{Diversity and statistics of the ProteinLMBench.}
The evaluation dataset consists of 944 questions with a median string length of 98-124 tokens, with a minimum of 46 and a maximum of 305 tokens. There are six answer choices, with an average string length of 77-114 tokens, with a minimum of 3 and a maximum of 368 tokens. The answers are well-balanced, with the correct answer being option 1 in 16.3\% of the cases, option 2 in 17.9\%, option 3 in 19.9\%, option 4 in 16.1\%, option 5 in 15.6\%, and option 6 in 14.0\% of the cases.

\paragraph{Safety during collection.}
In ensuring the safety and reliability of our proposed dataset, we have implemented rigorous measures and levels of filtering to address potential issues related to data integrity and quality. Given the inherent imperfections in structural and textual data, we acknowledge the importance of safeguarding against corrupted or erroneous information. Our dataset curation process includes thorough validation checks and quality control procedures to identify and rectify any anomalies that may compromise the integrity of the data. We are committed to transparency in our approach, encouraging users to report any instances of suspected data corruption or safety concerns through our designated channels. This proactive engagement with the research community is integral to maintaining the credibility of the dataset and fostering a collaborative environment for continual improvement. Additionally, our strategic categorization approach facilitates the identification and isolation of potential outliers, contributing to the overall safety and reliability of the dataset by providing users with a clear understanding of the distinct dimensions and sources within each segment. 
The ethical statement of this work is presented in Appendix Section \ref{ap_b2}.

\paragraph{Technical limitations.} \label{limitations}
Even when systematically collecting protein sequence data through scientific filtering and established resources, inherent technical limitations persist. Achieving perfection in structural data is rare, as experimental uncertainties stem from the limited resolution of techniques like X-ray crystallography or electron cryo-microscopy. 
\section{Experiment}
\label{ex}
\paragraph{Experiment setup.}
We conducted three different experiments: no training, supervised fine-tuning training (SFT), and self-supervised learning combined with supervised fine-tuning training (SSL-SFT) using \texttt{ProteinLMDataset}. For the no training experiment, we employed various large language models, including Falcon-7b \cite{f}, Qwen1.5-7B \cite{q}, Moonshot \cite{moon}, Mistral-7B-Instruct-v0.2 \cite{m}, Baichuan2-7B-Chat \cite{b}, Llama-2-7B-Chat-hf \cite{ll}, InternLM-Chat-20B, InternLM2-Chat-7B, and variations \cite{in}, ChatGLM3-6B \cite{ch}, Yi-6B-Chat \cite{yi}, GPT3.5-turbo, and GPT4.0-turbo \cite{gpt4}. For the SFT and SSL-SFT experiments, we utilized the InternLM2-7B model, naming them InternLM2-Protein-7B (w/o SSL) and InternLM2-Protein-7B, respectively. All the experiments were evaluated for accuracy using our proposed \texttt{ProteinLMBench}. For detailed training procedures and LLM architectures, please refer to Appendix C.

\begin{figure}[t]
    \centering    \includegraphics[width=0.88\textwidth]{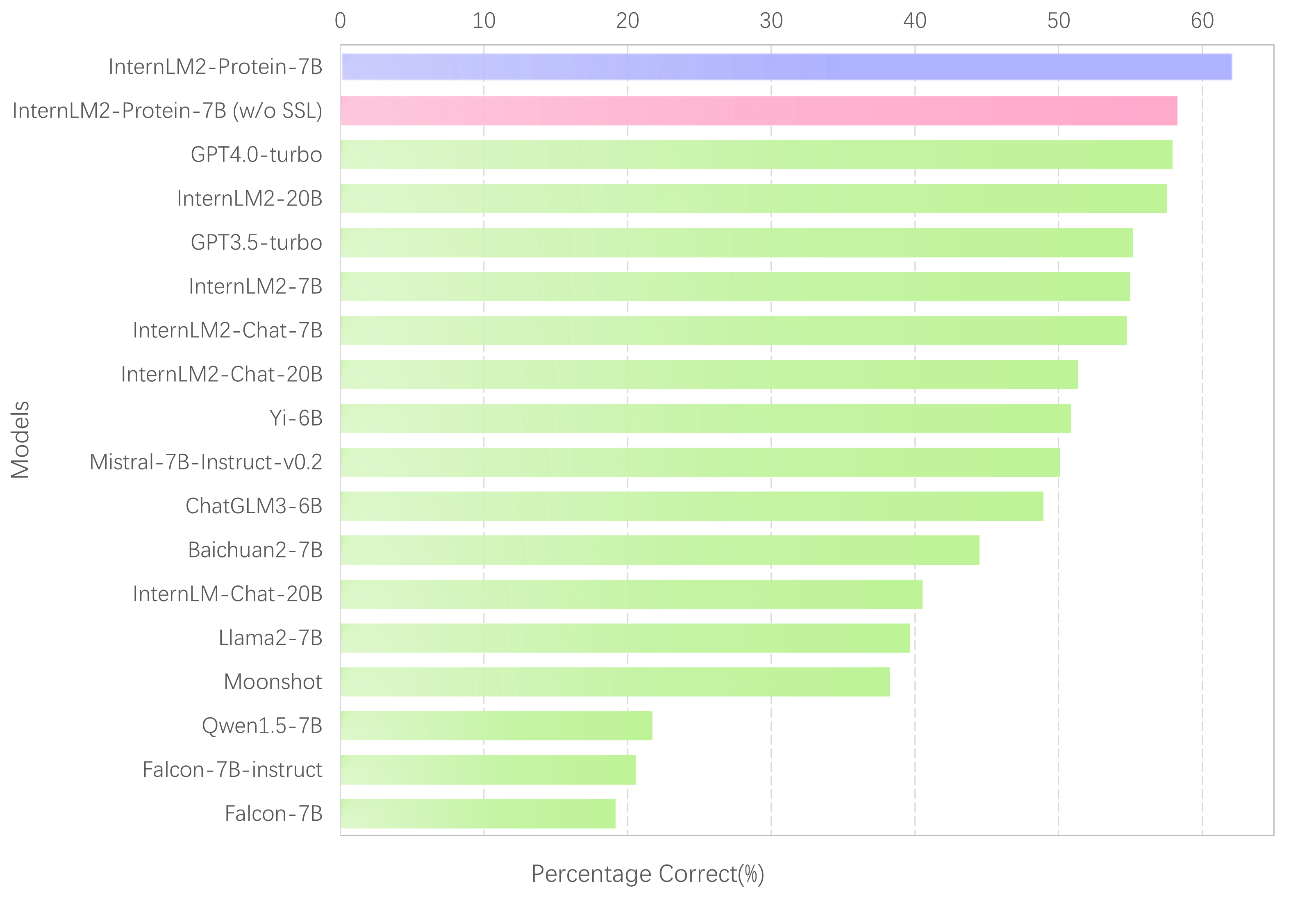}
    \caption{Model performence of LLMs on \texttt{ProteinLMBench}. The purple box plot represents the model performance combined training on \texttt{ProteinLMDataset} SSL and SFT datasets. The pink box plot represents the model performance obtained by only conducting SFT without SSL training. The green box plots represent the performance of other comparative models.}
    \label{sample}
\end{figure}

\paragraph{Experiment results.}
The results of all experiments are presented in Fig.~\ref{sample}. In the no-training experiment (depicted by green box plots), GPT-4.0-turbo achieved the highest accuracy with a correct rate of 57.94\%, followed by InternLM2-20B with a score of 57.52\%. Falcon-7B exhibited the poorest performance, achieving a correct rate of 19.17\%. In the supervised fine-tuning experiment (illustrated by pink box plots), InternLM2-7B delivered accuracy scores of 58.26\% (Fig. \ref{sample}; InternLM2-Protein-7B (w/o SSL)). This is an improvement from the no training experiment, where its accuracy was 54.98\%.
Notably, when we initially trained InternLM2-7B model on the self-supervised dataset and subsequently fine-tuned it (Fig. \ref{sample}; InternLM2-Protein-7B), the accuracy increased significantly to 62.18\%. This underscores the importance of the proposed \texttt{ProteinLMDataset} for more efficient and accurate protein understanding.

\section{Conclusion}
Our proposed dataset-benchmark tandem offers a comprehensive platform for training, fine-tuning, and evaluating language models on both Chinese-English text pairs and protein sequences, addressing limitations in existing datasets. The strategic classification approach ensures diversity and adaptability, tackling constraints within protein science and language model benchmarking. By curating a diverse collection from various sources, we reinforce the impartiality of our dataset. Incorporating both Chinese and English in detailing protein characteristics introduces a novel cross-lingual perspective, mitigating biases and enhancing multilingual understanding. Moreover, our proposed fine-tuning dataset enhances language models' capabilities in understanding, designing, and analyzing protein sequences. Experimentation within our \texttt{ProteinLMBench} benchmark validates the efficacy of our dataset. The InternLM2-7B language model achieved an accuracy of less than 55.00\% without further training on the \texttt{ProteinLMDataset}. When fine-tuned, its accuracy increased to 58.26\%. Notably, when InternLM2-7B was trained on both the self-supervised and fine-tuning datasets of \texttt{ProteinLMDataset}, it exhibited an exceptional improvement in accuracy, reaching 62.18\%. Our dataset and benchmark contribute to advancing artificial intelligence for biological sciences. It enables groundbreaking research and applications with the potential to transform diverse fields.

\bibliography{main.bib}

\begin{thebibliography}{37}
\providecommand{\natexlab}[1]{#1}
\providecommand{\url}[1]{\texttt{#1}}
\expandafter\ifx\csname urlstyle\endcsname\relax
  \providecommand{\doi}[1]{doi: #1}\else
  \providecommand{\doi}{doi: \begingroup \urlstyle{rm}\Url}\fi

\bibitem[01.AI et~al.(2024)01.AI, :, Young, Chen, Li, Huang, Zhang, Zhang, Li, Zhu, Chen, Chang, Yu, Liu, Liu, Yue, Yang, Yang, Yu, Xie, Huang, Hu, Ren, Niu, Nie, Xu, Liu, Wang, Cai, Gu, Liu, and Dai]{yi}
01.AI, :, A.~Young, B.~Chen, C.~Li, C.~Huang, G.~Zhang, G.~Zhang, H.~Li, J.~Zhu, J.~Chen, J.~Chang, K.~Yu, P.~Liu, Q.~Liu, S.~Yue, S.~Yang, S.~Yang, T.~Yu, W.~Xie, W.~Huang, X.~Hu, X.~Ren, X.~Niu, P.~Nie, Y.~Xu, Y.~Liu, Y.~Wang, Y.~Cai, Z.~Gu, Z.~Liu, and Z.~Dai.
\newblock {Yi: Open Foundation Models by 01.AI}.
\newblock \emph{arXiv preprint arXiv:2403.04652}, 2024.

\bibitem[Almazrouei et~al.(2023)Almazrouei, Alobeidli, Alshamsi, Cappelli, Cojocaru, Debbah, Étienne Goffinet, Hesslow, Launay, Malartic, Mazzotta, Noune, Pannier, and Penedo]{f}
E.~Almazrouei, H.~Alobeidli, A.~Alshamsi, A.~Cappelli, R.~Cojocaru, M.~Debbah, Étienne Goffinet, D.~Hesslow, J.~Launay, Q.~Malartic, D.~Mazzotta, B.~Noune, B.~Pannier, and G.~Penedo.
\newblock The {Falcon Series of Open Language Models}.
\newblock \emph{arXiv preprint arXiv:2311.16867}, 2023.

\bibitem[Bai et~al.(2023)Bai, Bai, Chu, Cui, Dang, Deng, Fan, Ge, Han, Huang, Hui, Ji, Li, Lin, Lin, Liu, Liu, Lu, Lu, Ma, Men, Ren, Ren, Tan, Tan, Tu, Wang, Wang, Wang, Wu, Xu, Xu, Yang, Yang, Yang, Yang, Yao, Yu, Yuan, Yuan, Zhang, Zhang, Zhang, Zhang, Zhou, Zhou, Zhou, and Zhu]{q}
J.~Bai, S.~Bai, Y.~Chu, Z.~Cui, K.~Dang, X.~Deng, Y.~Fan, W.~Ge, Y.~Han, F.~Huang, B.~Hui, L.~Ji, M.~Li, J.~Lin, R.~Lin, D.~Liu, G.~Liu, C.~Lu, K.~Lu, J.~Ma, R.~Men, X.~Ren, X.~Ren, C.~Tan, S.~Tan, J.~Tu, P.~Wang, S.~Wang, W.~Wang, S.~Wu, B.~Xu, J.~Xu, A.~Yang, H.~Yang, J.~Yang, S.~Yang, Y.~Yao, B.~Yu, H.~Yuan, Z.~Yuan, J.~Zhang, X.~Zhang, Y.~Zhang, Z.~Zhang, C.~Zhou, J.~Zhou, X.~Zhou, and T.~Zhu.
\newblock Qwen technical report.
\newblock \emph{arXiv preprint arXiv:2309.16609}, 2023.

\bibitem[Baichuan(2023)]{b}
Baichuan.
\newblock Baichuan 2: Open large-scale language models.
\newblock \emph{arXiv preprint arXiv:2309.10305}, 2023.
\newblock URL \url{https://arxiv.org/abs/2309.10305}.

\bibitem[Blum et~al.(2020)Blum, Chang, Chuguransky, Grego, Kandasaamy, Mitchell, Nuka, Paysan-Lafosse, Qureshi, Raj, Richardson, Salazar, Williams, Bork, Bridge, Gough, Haft, Letunic, Marchler-Bauer, Mi, Natale, Necci, Orengo, Pandurangan, Rivoire, Sigrist, Sillitoe, Thanki, Thomas, Tosatto, Wu, Bateman, and Finn]{18}
M.~Blum, H.-Y. Chang, S.~Chuguransky, T.~Grego, S.~Kandasaamy, A.~Mitchell, G.~Nuka, T.~Paysan-Lafosse, M.~Qureshi, S.~Raj, L.~Richardson, G.~A. Salazar, L.~Williams, P.~Bork, A.~Bridge, J.~Gough, D.~H. Haft, I.~Letunic, A.~Marchler-Bauer, H.~Mi, D.~A. Natale, M.~Necci, C.~A. Orengo, A.~P. Pandurangan, C.~Rivoire, C.~J.~A. Sigrist, I.~Sillitoe, N.~Thanki, P.~D. Thomas, S.~C.~E. Tosatto, C.~H. Wu, A.~Bateman, and R.~D. Finn.
\newblock {The InterPro protein families and domains database: 20 years on}.
\newblock \emph{Nucleic Acids Research}, 49\penalty0 (D1):\penalty0 D344--D354, 11 2020.
\newblock ISSN 0305-1048.
\newblock \doi{10.1093/nar/gkaa977}.
\newblock URL \url{https://doi.org/10.1093/nar/gkaa977}.

\bibitem[Burley et~al.(2020)Burley, Bhikadiya, Bi, Bittrich, Chen, Crichlow, Christie, Dalenberg, Di~Costanzo, Duarte, Dutta, Feng, Ganesan, Goodsell, Ghosh, Green, Guranović, Guzenko, Hudson, Lawson, Liang, Lowe, Namkoong, Peisach, Persikova, Randle, Rose, Rose, Sali, Segura, Sekharan, Shao, Tao, Voigt, Westbrook, Young, Zardecki, and Zhuravleva]{11}
S.~K. Burley, C.~Bhikadiya, C.~Bi, S.~Bittrich, L.~Chen, G.~V. Crichlow, C.~H. Christie, K.~Dalenberg, L.~Di~Costanzo, J.~M. Duarte, S.~Dutta, Z.~Feng, S.~Ganesan, D.~S. Goodsell, S.~Ghosh, R.~K. Green, V.~Guranović, D.~Guzenko, B.~P. Hudson, C.~Lawson, Y.~Liang, R.~Lowe, H.~Namkoong, E.~Peisach, I.~Persikova, C.~Randle, A.~Rose, Y.~Rose, A.~Sali, J.~Segura, M.~Sekharan, C.~Shao, Y.-P. Tao, M.~Voigt, J.~Westbrook, J.~Y. Young, C.~Zardecki, and M.~Zhuravleva.
\newblock {RCSB Protein Data Bank: powerful new tools for exploring 3D structures of biological macromolecules for basic and applied research and education in fundamental biology, biomedicine, biotechnology, bioengineering and energy sciences}.
\newblock \emph{Nucleic Acids Research}, 49\penalty0 (D1):\penalty0 D437--D451, 11 2020.
\newblock ISSN 0305-1048.
\newblock \doi{10.1093/nar/gkaa1038}.
\newblock URL \url{https://doi.org/10.1093/nar/gkaa1038}.

\bibitem[Cai et~al.(2024)Cai, Cao, Chen, Chen, Chen, Chen, Chen, Chen, Chen, Chu, Dong, Duan, Fan, Fei, Gao, Ge, Gu, Gu, Gui, Guo, Guo, He, Hu, Huang, Jiang, Jiao, Jin, Lei, Li, Li, Li, Li, Li, Li, Liu, Liu, Hong, Liu, Liu, Liu, Lv, Lv, Lv, Ma, Ma, Ma, Ning, Ouyang, Qiu, Qu, Shang, Shao, Song, Song, Sui, Sun, Sun, Tang, Wang, Wang, Wang, Wang, Wang, Wang, Wang, Wei, Weng, Wu, Xiong, Xu, Xu, Yan, Yan, Yang, Ye, Ying, Yu, Yu, Zang, Zhang, Zhang, Zhang, Zhang, Zhang, Zhang, Zhang, Zhang, Zhang, Zhang, Zhang, Zhao, Zhao, Zhao, Zhou, Zhou, Zhuo, Zou, Qiu, Qiao, and Lin]{in}
Z.~Cai, M.~Cao, H.~Chen, K.~Chen, K.~Chen, X.~Chen, X.~Chen, Z.~Chen, Z.~Chen, P.~Chu, X.~Dong, H.~Duan, Q.~Fan, Z.~Fei, Y.~Gao, J.~Ge, C.~Gu, Y.~Gu, T.~Gui, A.~Guo, Q.~Guo, C.~He, Y.~Hu, T.~Huang, T.~Jiang, P.~Jiao, Z.~Jin, Z.~Lei, J.~Li, J.~Li, L.~Li, S.~Li, W.~Li, Y.~Li, H.~Liu, J.~Liu, J.~Hong, K.~Liu, K.~Liu, X.~Liu, C.~Lv, H.~Lv, K.~Lv, L.~Ma, R.~Ma, Z.~Ma, W.~Ning, L.~Ouyang, J.~Qiu, Y.~Qu, F.~Shang, Y.~Shao, D.~Song, Z.~Song, Z.~Sui, P.~Sun, Y.~Sun, H.~Tang, B.~Wang, G.~Wang, J.~Wang, J.~Wang, R.~Wang, Y.~Wang, Z.~Wang, X.~Wei, Q.~Weng, F.~Wu, Y.~Xiong, C.~Xu, R.~Xu, H.~Yan, Y.~Yan, X.~Yang, H.~Ye, H.~Ying, J.~Yu, J.~Yu, Y.~Zang, C.~Zhang, L.~Zhang, P.~Zhang, P.~Zhang, R.~Zhang, S.~Zhang, S.~Zhang, W.~Zhang, W.~Zhang, X.~Zhang, X.~Zhang, H.~Zhao, Q.~Zhao, X.~Zhao, F.~Zhou, Z.~Zhou, J.~Zhuo, Y.~Zou, X.~Qiu, Y.~Qiao, and D.~Lin.
\newblock Internlm2 technical report.
\newblock \emph{arXiv preprint arXiv:2403.17297}, 2024.

\bibitem[Consortium(2022)]{14}
T.~U. Consortium.
\newblock {UniProt: the Universal Protein Knowledgebase in 2023}.
\newblock \emph{Nucleic Acids Research}, 51\penalty0 (D1):\penalty0 D523--D531, 11 2022.
\newblock ISSN 0305-1048.
\newblock \doi{10.1093/nar/gkac1052}.
\newblock URL \url{https://doi.org/10.1093/nar/gkac1052}.

\bibitem[Du et~al.(2022)Du, Qian, Liu, Ding, Qiu, Yang, and Tang]{ch}
Z.~Du, Y.~Qian, X.~Liu, M.~Ding, J.~Qiu, Z.~Yang, and J.~Tang.
\newblock {GLM: General Language Model Pretraining with Autoregressive Blank Infilling}.
\newblock In \emph{ACL (Volume 1: Long Papers)}, pages 320--335, 2022.

\bibitem[Fang et~al.(2024)Fang, Liang, Zhang, Liu, Huang, Chen, Fan, and Chen]{icrl24}
Y.~Fang, X.~Liang, N.~Zhang, K.~Liu, R.~Huang, Z.~Chen, X.~Fan, and H.~Chen.
\newblock {Mol-Instructions: A Large-Scale Biomolecular Instruction Dataset for Large Language Models}.
\newblock In \emph{ICLR}, 2024.
\newblock URL \url{https://openreview.net/forum?id=Tlsdsb6l9n}.

\bibitem[Gao et~al.(2024)]{rag}
Y.~Gao et~al.
\newblock Retrieval-augmented generation for large language models: A survey.
\newblock \emph{arXiv preprint arXiv:2312.10997}, 2024.

\bibitem[Hendrycks et~al.(2021)Hendrycks, Burns, Basart, Zou, Mazeika, Song, and Steinhardt]{4}
D.~Hendrycks, C.~Burns, S.~Basart, A.~Zou, M.~Mazeika, D.~Song, and J.~Steinhardt.
\newblock Measuring massive multitask language understanding.
\newblock In \emph{ICLR}, 2021.
\newblock URL \url{https://openreview.net/forum?id=d7KBjmI3GmQ}.

\bibitem[Huang et~al.(2023)Huang, Bai, Zhu, Zhang, Zhang, Su, Liu, Lv, Zhang, lei, Fu, Sun, and He]{7}
Y.~Huang, Y.~Bai, Z.~Zhu, J.~Zhang, J.~Zhang, T.~Su, J.~Liu, C.~Lv, Y.~Zhang, j.~lei, Y.~Fu, M.~Sun, and J.~He.
\newblock {C-Eval: A Multi-Level Multi-Discipline Chinese Evaluation Suite for Foundation Models}.
\newblock In \emph{NeurIPS}, volume~36, pages 62991--63010, 2023.

\bibitem[Jiang et~al.(2023)Jiang, Sablayrolles, Mensch, Bamford, Chaplot, de~las Casas, Bressand, Lengyel, Lample, Saulnier, Lavaud, Lachaux, Stock, Scao, Lavril, Wang, Lacroix, and Sayed]{m}
A.~Q. Jiang, A.~Sablayrolles, A.~Mensch, C.~Bamford, D.~S. Chaplot, D.~de~las Casas, F.~Bressand, G.~Lengyel, G.~Lample, L.~Saulnier, L.~R. Lavaud, M.-A. Lachaux, P.~Stock, T.~L. Scao, T.~Lavril, T.~Wang, T.~Lacroix, and W.~E. Sayed.
\newblock {Mistral 7B}.
\newblock \emph{arXiv preprint arXiv:2310.06825}, 2023.

\bibitem[Kanehisa et~al.(2022)Kanehisa, Furumichi, Sato, Kawashima, and Ishiguro-Watanabe]{16}
M.~Kanehisa, M.~Furumichi, Y.~Sato, M.~Kawashima, and M.~Ishiguro-Watanabe.
\newblock {KEGG for taxonomy-based analysis of pathways and genomes}.
\newblock \emph{Nucleic Acids Research}, 51\penalty0 (D1):\penalty0 D587--D592, 10 2022.
\newblock ISSN 0305-1048.
\newblock \doi{10.1093/nar/gkac963}.
\newblock URL \url{https://doi.org/10.1093/nar/gkac963}.

\bibitem[Kinch et~al.(2021)Kinch, Schaeffer, Kryshtafovych, and Grishin]{13}
L.~Kinch, R.~Schaeffer, A.~Kryshtafovych, and N.~Grishin.
\newblock {Target classification in the 14th round of the critical assessment of protein structure prediction (CASP14)}.
\newblock \emph{Proteins: Structure, Function and Bioinformatics}, 89\penalty0 (12):\penalty0 1618--1632, Dec. 2021.
\newblock ISSN 0887-3585.
\newblock \doi{10.1002/prot.26202}.

\bibitem[Liang et~al.(2023)Liang, Bommasani, Lee, Tsipras, Soylu, Yasunaga, Zhang, Narayanan, Wu, Kumar, Newman, Yuan, Yan, Zhang, Cosgrove, Manning, Re, Acosta-Navas, Hudson, Zelikman, Durmus, Ladhak, Rong, Ren, Yao, WANG, Santhanam, Orr, Zheng, Yuksekgonul, Suzgun, Kim, Guha, Chatterji, Khattab, Henderson, Huang, Chi, Xie, Santurkar, Ganguli, Hashimoto, Icard, Zhang, Chaudhary, Wang, Li, Mai, Zhang, and Koreeda]{6}
P.~Liang, R.~Bommasani, T.~Lee, D.~Tsipras, D.~Soylu, M.~Yasunaga, Y.~Zhang, D.~Narayanan, Y.~Wu, A.~Kumar, B.~Newman, B.~Yuan, B.~Yan, C.~Zhang, C.~A. Cosgrove, C.~D. Manning, C.~Re, D.~Acosta-Navas, D.~A. Hudson, E.~Zelikman, E.~Durmus, F.~Ladhak, F.~Rong, H.~Ren, H.~Yao, J.~WANG, K.~Santhanam, L.~Orr, L.~Zheng, M.~Yuksekgonul, M.~Suzgun, N.~Kim, N.~Guha, N.~S. Chatterji, O.~Khattab, P.~Henderson, Q.~Huang, R.~A. Chi, S.~M. Xie, S.~Santurkar, S.~Ganguli, T.~Hashimoto, T.~Icard, T.~Zhang, V.~Chaudhary, W.~Wang, X.~Li, Y.~Mai, Y.~Zhang, and Y.~Koreeda.
\newblock {Holistic Evaluation of Language Models}.
\newblock \emph{Transactions on Machine Learning Research}, 2023.
\newblock ISSN 2835-8856.
\newblock URL \url{https://openreview.net/forum?id=iO4LZibEqW}.
\newblock Featured Certification, Expert Certification.

\bibitem[Narayan et~al.(2018)Narayan, Cohen, and Lapata]{3}
S.~Narayan, S.~B. Cohen, and M.~Lapata.
\newblock Don{'}t give me the details, just the summary! topic-aware convolutional neural networks for extreme summarization.
\newblock In \emph{EMNLP}, pages 1797--1807, 2018.
\newblock \doi{10.18653/v1/D18-1206}.
\newblock URL \url{https://aclanthology.org/D18-1206}.

\bibitem[O'Leary et~al.(2015)O'Leary, Wright, Brister, Ciufo, Haddad, McVeigh, Rajput, Robbertse, Smith-White, Ako-Adjei, Astashyn, Badretdin, Bao, Blinkova, Brover, Chetvernin, Choi, Cox, Ermolaeva, Farrell, Goldfarb, Gupta, Haft, Hatcher, Hlavina, Joardar, Kodali, Li, Maglott, Masterson, McGarvey, Murphy, O'Neill, Pujar, Rangwala, Rausch, Riddick, Schoch, Shkeda, Storz, Sun, Thibaud-Nissen, Tolstoy, Tully, Vatsan, Wallin, Webb, Wu, Landrum, Kimchi, Tatusova, DiCuccio, Kitts, Murphy, and Pruitt]{15}
N.~A. O'Leary, M.~W. Wright, J.~R. Brister, S.~Ciufo, D.~Haddad, R.~McVeigh, B.~Rajput, B.~Robbertse, B.~Smith-White, D.~Ako-Adjei, A.~Astashyn, A.~Badretdin, Y.~Bao, O.~Blinkova, V.~Brover, V.~Chetvernin, J.~Choi, E.~Cox, O.~Ermolaeva, C.~M. Farrell, T.~Goldfarb, T.~Gupta, D.~Haft, E.~Hatcher, W.~Hlavina, V.~S. Joardar, V.~K. Kodali, W.~Li, D.~Maglott, P.~Masterson, K.~M. McGarvey, M.~R. Murphy, K.~O'Neill, S.~Pujar, S.~H. Rangwala, D.~Rausch, L.~D. Riddick, C.~Schoch, A.~Shkeda, S.~S. Storz, H.~Sun, F.~Thibaud-Nissen, I.~Tolstoy, R.~E. Tully, A.~R. Vatsan, C.~Wallin, D.~Webb, W.~Wu, M.~J. Landrum, A.~Kimchi, T.~Tatusova, M.~DiCuccio, P.~Kitts, T.~D. Murphy, and K.~D. Pruitt.
\newblock {Reference sequence (RefSeq) database at NCBI: current status, taxonomic expansion, and functional annotation}.
\newblock \emph{Nucleic Acids Research}, 44\penalty0 (D1):\penalty0 D733--D745, 11 2015.
\newblock ISSN 0305-1048.
\newblock \doi{10.1093/nar/gkv1189}.
\newblock URL \url{https://doi.org/10.1093/nar/gkv1189}.

\bibitem[OpenAI et~al.(2024)]{gpt4}
OpenAI et~al.
\newblock {GPT-4 Technical Report}.
\newblock \emph{arxiv preprint arXiv:2303.08774}, 2024.

\bibitem[Orengo et~al.(1997)Orengo, Michie, Jones, Jones, Swindells, and Thornton]{10}
C.~Orengo, A.~Michie, S.~Jones, D.~Jones, M.~Swindells, and J.~Thornton.
\newblock {CATH – a hierarchic classification of protein domain structures}.
\newblock \emph{Structure}, 5\penalty0 (8):\penalty0 1093--1109, 1997.
\newblock ISSN 0969-2126.
\newblock \doi{https://doi.org/10.1016/S0969-2126(97)00260-8}.
\newblock URL \url{https://www.sciencedirect.com/science/article/pii/S0969212697002608}.

\bibitem[Rajpurkar et~al.(2018)Rajpurkar, Jia, and Liang]{2}
P.~Rajpurkar, R.~Jia, and P.~Liang.
\newblock {Know What You Don't Know: Unanswerable Questions for SQuAD}.
\newblock In \emph{ACL}, 2018.

\bibitem[Rives et~al.(2019)Rives, Meier, Sercu, Goyal, Lin, Liu, Guo, Ott, Zitnick, Ma, and Fergus]{rives2019biological}
A.~Rives, J.~Meier, T.~Sercu, S.~Goyal, Z.~Lin, J.~Liu, D.~Guo, M.~Ott, C.~L. Zitnick, J.~Ma, and R.~Fergus.
\newblock Biological structure and function emerge from scaling unsupervised learning to 250 million protein sequences.
\newblock \emph{PNAS}, 2019.
\newblock \doi{10.1101/622803}.
\newblock URL \url{https://www.biorxiv.org/content/10.1101/622803v4}.

\bibitem[Sayers et~al.(2021)Sayers, Bolton, Brister, Canese, Chan, Comeau, Connor, Funk, Kelly, Kim, Madej, Marchler-Bauer, Lanczycki, Lathrop, Lu, Thibaud-Nissen, Murphy, Phan, Skripchenko, Tse, Wang, Williams, Trawick, Pruitt, and Sherry]{ncbi}
E.~W. Sayers, E.~E. Bolton, J.~R. Brister, K.~Canese, J.~Chan, D.~C. Comeau, R.~Connor, K.~Funk, C.~Kelly, S.~Kim, T.~Madej, A.~Marchler-Bauer, C.~Lanczycki, S.~Lathrop, Z.~Lu, F.~Thibaud-Nissen, T.~Murphy, L.~Phan, Y.~Skripchenko, T.~Tse, J.~Wang, R.~Williams, B.~W. Trawick, K.~D. Pruitt, and S.~T. Sherry.
\newblock {Database resources of the national center for biotechnology information}.
\newblock \emph{Nucleic Acids Research}, 50\penalty0 (D1):\penalty0 D20--D26, 12 2021.
\newblock ISSN 0305-1048.
\newblock \doi{10.1093/nar/gkab1112}.
\newblock URL \url{https://doi.org/10.1093/nar/gkab1112}.

\bibitem[Schuhmann et~al.(2022)Schuhmann, Beaumont, Vencu, Gordon, Wightman, Cherti, Coombes, Katta, Mullis, Wortsman, et~al.]{app}
C.~Schuhmann, R.~Beaumont, R.~Vencu, C.~Gordon, R.~Wightman, M.~Cherti, T.~Coombes, A.~Katta, C.~Mullis, M.~Wortsman, et~al.
\newblock {LAION-5B:} an open large-scale dataset for training next generation image-text models.
\newblock In \emph{NeurIPS}, volume~35, pages 25278--25294, 2022.

\bibitem[Senior et~al.(2019)Senior, Evans, Jumper, Kirkpatrick, Sifre, Green, Qin, Žídek, Nelson, Bridgland, Penedones, Petersen, Simonyan, Crossan, Kohli, Jones, Silver, Kavukcuoglu, and Hassabis]{12}
A.~W. Senior, R.~Evans, J.~Jumper, J.~Kirkpatrick, L.~Sifre, T.~Green, C.~Qin, A.~Žídek, A.~W.~R. Nelson, A.~Bridgland, H.~Penedones, S.~Petersen, K.~Simonyan, S.~Crossan, P.~Kohli, D.~T. Jones, D.~Silver, K.~Kavukcuoglu, and D.~Hassabis.
\newblock {Protein structure prediction using multiple deep neural networks in the 13th Critical Assessment of Protein Structure Prediction (CASP13)}.
\newblock \emph{Proteins: Structure, Function, and Bioinformatics}, 87\penalty0 (12):\penalty0 1141--1148, 2019.
\newblock \doi{https://doi.org/10.1002/prot.25834}.
\newblock URL \url{https://onlinelibrary.wiley.com/doi/abs/10.1002/prot.25834}.

\bibitem[Srivastava et~al.(2023)Srivastava, Rastogi, Rao, Shoeb, Abid, Fisch, Brown, Santoro, Gupta, Garriga-Alonso, Kluska, Lewkowycz, Agarwal, Power, Ray, Warstadt, Kocurek, Safaya, Tazarv, Xiang, Parrish, Nie, Hussain, Askell, Dsouza, Slone, Rahane, Iyer, Andreassen, Madotto, Santilli, Stuhlm{\"u}ller, Dai, La, Lampinen, Zou, Jiang, Chen, Vuong, Gupta, Gottardi, Norelli, Venkatesh, Gholamidavoodi, Tabassum, Menezes, Kirubarajan, Mullokandov, Sabharwal, Herrick, Efrat, Erdem, Karaka{\c{s}}, Roberts, Loe, Zoph, Bojanowski, {\"O}zyurt, Hedayatnia, Neyshabur, Inden, Stein, Ekmekci, Lin, Howald, Orinion, Diao, Dour, Stinson, Argueta, Ferri, Singh, Rathkopf, Meng, Baral, Wu, Callison-Burch, Waites, Voigt, Manning, Potts, Ramirez, Rivera, Siro, Raffel, Ashcraft, Garbacea, Sileo, Garrette, Hendrycks, Kilman, Roth, Freeman, Khashabi, Levy, Gonz{\'a}lez, Perszyk, Hernandez, Chen, Ippolito, Gilboa, Dohan, Drakard, Jurgens, Datta, Ganguli, Emelin, Kleyko, Yuret, Chen, Tam, Hupkes, Misra, Buzan, Mollo, Yang, Lee,
  Schrader, Shutova, Cubuk, Segal, Hagerman, Barnes, Donoway, Pavlick, Rodol{\`a}, Lam, Chu, Tang, Erdem, Chang, Chi, Dyer, Jerzak, Kim, Manyasi, Zheltonozhskii, Xia, Siar, Mart{\'\i}nez-Plumed, Happ{\'e}, Chollet, Rong, Mishra, Winata, de~Melo, Kruszewski, Parascandolo, Mariani, Wang, Jaimovitch-Lopez, Betz, Gur-Ari, Galijasevic, Kim, Rashkin, Hajishirzi, Mehta, Bogar, Shevlin, Schuetze, Yakura, Zhang, Wong, Ng, Noble, Jumelet, Geissinger, Kernion, Hilton, Lee, Fisac, Simon, Koppel, Zheng, Zou, Kocon, Thompson, Wingfield, Kaplan, Radom, Sohl-Dickstein, Phang, Wei, Yosinski, Novikova, Bosscher, Marsh, Kim, Taal, Engel, Alabi, Xu, Song, Tang, Waweru, Burden, Miller, Balis, Batchelder, Berant, Frohberg, Rozen, Hernandez-Orallo, Boudeman, Guerr, Jones, Tenenbaum, Rule, Chua, Kanclerz, Livescu, Krauth, Gopalakrishnan, Ignatyeva, Markert, Dhole, Gimpel, Omondi, Mathewson, Chiafullo, Shkaruta, Shridhar, McDonell, Richardson, Reynolds, Gao, Zhang, Dugan, Qin, Contreras-Ochando, Morency, Moschella, Lam, Noble,
  Schmidt, He, Oliveros-Col{\'o}n, Metz, Senel, Bosma, Sap, Hoeve, Farooqi, Faruqui, Mazeika, Baturan, Marelli, Maru, Ramirez-Quintana, Tolkiehn, Giulianelli, Lewis, Potthast, Leavitt, Hagen, Schubert, Baitemirova, Arnaud, McElrath, Yee, Cohen, Gu, Ivanitskiy, Starritt, Strube, Sw{\k{e}}drowski, Bevilacqua, Yasunaga, Kale, Cain, Xu, Suzgun, Walker, Tiwari, Bansal, Aminnaseri, Geva, Gheini, T, Peng, Chi, Lee, Krakover, Cameron, Roberts, Doiron, Martinez, Nangia, Deckers, Muennighoff, Keskar, Iyer, Constant, Fiedel, Wen, Zhang, Agha, Elbaghdadi, Levy, Evans, Casares, Doshi, Fung, Liang, Vicol, Alipoormolabashi, Liao, Liang, Chang, Eckersley, Htut, Hwang, Mi{\l}kowski, Patil, Pezeshkpour, Oli, Mei, Lyu, Chen, Banjade, Rudolph, Gabriel, Habacker, Risco, Milli{\`e}re, Garg, Barnes, Saurous, Arakawa, Raymaekers, Frank, Sikand, Novak, Sitelew, Bras, Liu, Jacobs, Zhang, Salakhutdinov, Chi, Lee, Stovall, Teehan, Yang, Singh, Mohammad, Anand, Dillavou, Shleifer, Wiseman, Gruetter, Bowman, Schoenholz, Han, Kwatra, Rous,
  Ghazarian, Ghosh, Casey, Bischoff, Gehrmann, Schuster, Sadeghi, Hamdan, Zhou, Srivastava, Shi, Singh, Asaadi, Gu, Pachchigar, Toshniwal, Upadhyay, Debnath, Shakeri, Thormeyer, Melzi, Reddy, Makini, Lee, Torene, Hatwar, Dehaene, Divic, Ermon, Biderman, Lin, Prasad, Piantadosi, Shieber, Misherghi, Kiritchenko, Mishra, Linzen, Schuster, Li, Yu, Ali, Hashimoto, Wu, Desbordes, Rothschild, Phan, Wang, Nkinyili, Schick, Kornev, Tunduny, Gerstenberg, Chang, Neeraj, Khot, Shultz, Shaham, Misra, Demberg, Nyamai, Raunak, Ramasesh, vinay~uday prabhu, Padmakumar, Srikumar, Fedus, Saunders, Zhang, Vossen, Ren, Tong, Zhao, Wu, Shen, Yaghoobzadeh, Lakretz, Song, Bahri, Choi, Yang, Hao, Chen, Belinkov, Hou, Hou, Bai, Seid, Zhao, Wang, Wang, Wang, and Wu]{5}
A.~Srivastava, A.~Rastogi, A.~Rao, A.~A.~M. Shoeb, A.~Abid, A.~Fisch, A.~R. Brown, A.~Santoro, A.~Gupta, A.~Garriga-Alonso, A.~Kluska, A.~Lewkowycz, A.~Agarwal, A.~Power, A.~Ray, A.~Warstadt, A.~W. Kocurek, A.~Safaya, A.~Tazarv, A.~Xiang, A.~Parrish, A.~Nie, A.~Hussain, A.~Askell, A.~Dsouza, A.~Slone, A.~Rahane, A.~S. Iyer, A.~J. Andreassen, A.~Madotto, A.~Santilli, A.~Stuhlm{\"u}ller, A.~M. Dai, A.~La, A.~Lampinen, A.~Zou, A.~Jiang, A.~Chen, A.~Vuong, A.~Gupta, A.~Gottardi, A.~Norelli, A.~Venkatesh, A.~Gholamidavoodi, A.~Tabassum, A.~Menezes, A.~Kirubarajan, A.~Mullokandov, A.~Sabharwal, A.~Herrick, A.~Efrat, A.~Erdem, A.~Karaka{\c{s}}, B.~R. Roberts, B.~S. Loe, B.~Zoph, B.~Bojanowski, B.~{\"O}zyurt, B.~Hedayatnia, B.~Neyshabur, B.~Inden, B.~Stein, B.~Ekmekci, B.~Y. Lin, B.~Howald, B.~Orinion, C.~Diao, C.~Dour, C.~Stinson, C.~Argueta, C.~Ferri, C.~Singh, C.~Rathkopf, C.~Meng, C.~Baral, C.~Wu, C.~Callison-Burch, C.~Waites, C.~Voigt, C.~D. Manning, C.~Potts, C.~Ramirez, C.~E. Rivera, C.~Siro, C.~Raffel,
  C.~Ashcraft, C.~Garbacea, D.~Sileo, D.~Garrette, D.~Hendrycks, D.~Kilman, D.~Roth, C.~D. Freeman, D.~Khashabi, D.~Levy, D.~M. Gonz{\'a}lez, D.~Perszyk, D.~Hernandez, D.~Chen, D.~Ippolito, D.~Gilboa, D.~Dohan, D.~Drakard, D.~Jurgens, D.~Datta, D.~Ganguli, D.~Emelin, D.~Kleyko, D.~Yuret, D.~Chen, D.~Tam, D.~Hupkes, D.~Misra, D.~Buzan, D.~C. Mollo, D.~Yang, D.-H. Lee, D.~Schrader, E.~Shutova, E.~D. Cubuk, E.~Segal, E.~Hagerman, E.~Barnes, E.~Donoway, E.~Pavlick, E.~Rodol{\`a}, E.~Lam, E.~Chu, E.~Tang, E.~Erdem, E.~Chang, E.~A. Chi, E.~Dyer, E.~Jerzak, E.~Kim, E.~E. Manyasi, E.~Zheltonozhskii, F.~Xia, F.~Siar, F.~Mart{\'\i}nez-Plumed, F.~Happ{\'e}, F.~Chollet, F.~Rong, G.~Mishra, G.~I. Winata, G.~de~Melo, G.~Kruszewski, G.~Parascandolo, G.~Mariani, G.~X. Wang, G.~Jaimovitch-Lopez, G.~Betz, G.~Gur-Ari, H.~Galijasevic, H.~Kim, H.~Rashkin, H.~Hajishirzi, H.~Mehta, H.~Bogar, H.~F.~A. Shevlin, H.~Schuetze, H.~Yakura, H.~Zhang, H.~M. Wong, I.~Ng, I.~Noble, J.~Jumelet, J.~Geissinger, J.~Kernion, J.~Hilton, J.~Lee,
  J.~F. Fisac, J.~B. Simon, J.~Koppel, J.~Zheng, J.~Zou, J.~Kocon, J.~Thompson, J.~Wingfield, J.~Kaplan, J.~Radom, J.~Sohl-Dickstein, J.~Phang, J.~Wei, J.~Yosinski, J.~Novikova, J.~Bosscher, J.~Marsh, J.~Kim, J.~Taal, J.~Engel, J.~Alabi, J.~Xu, J.~Song, J.~Tang, J.~Waweru, J.~Burden, J.~Miller, J.~U. Balis, J.~Batchelder, J.~Berant, J.~Frohberg, J.~Rozen, J.~Hernandez-Orallo, J.~Boudeman, J.~Guerr, J.~Jones, J.~B. Tenenbaum, J.~S. Rule, J.~Chua, K.~Kanclerz, K.~Livescu, K.~Krauth, K.~Gopalakrishnan, K.~Ignatyeva, K.~Markert, K.~Dhole, K.~Gimpel, K.~Omondi, K.~W. Mathewson, K.~Chiafullo, K.~Shkaruta, K.~Shridhar, K.~McDonell, K.~Richardson, L.~Reynolds, L.~Gao, L.~Zhang, L.~Dugan, L.~Qin, L.~Contreras-Ochando, L.-P. Morency, L.~Moschella, L.~Lam, L.~Noble, L.~Schmidt, L.~He, L.~Oliveros-Col{\'o}n, L.~Metz, L.~K. Senel, M.~Bosma, M.~Sap, M.~T. Hoeve, M.~Farooqi, M.~Faruqui, M.~Mazeika, M.~Baturan, M.~Marelli, M.~Maru, M.~J. Ramirez-Quintana, M.~Tolkiehn, M.~Giulianelli, M.~Lewis, M.~Potthast, M.~L. Leavitt,
  M.~Hagen, M.~Schubert, M.~O. Baitemirova, M.~Arnaud, M.~McElrath, M.~A. Yee, M.~Cohen, M.~Gu, M.~Ivanitskiy, M.~Starritt, M.~Strube, M.~Sw{\k{e}}drowski, M.~Bevilacqua, M.~Yasunaga, M.~Kale, M.~Cain, M.~Xu, M.~Suzgun, M.~Walker, M.~Tiwari, M.~Bansal, M.~Aminnaseri, M.~Geva, M.~Gheini, M.~V. T, N.~Peng, N.~A. Chi, N.~Lee, N.~G.-A. Krakover, N.~Cameron, N.~Roberts, N.~Doiron, N.~Martinez, N.~Nangia, N.~Deckers, N.~Muennighoff, N.~S. Keskar, N.~S. Iyer, N.~Constant, N.~Fiedel, N.~Wen, O.~Zhang, O.~Agha, O.~Elbaghdadi, O.~Levy, O.~Evans, P.~A.~M. Casares, P.~Doshi, P.~Fung, P.~P. Liang, P.~Vicol, P.~Alipoormolabashi, P.~Liao, P.~Liang, P.~W. Chang, P.~Eckersley, P.~M. Htut, P.~Hwang, P.~Mi{\l}kowski, P.~Patil, P.~Pezeshkpour, P.~Oli, Q.~Mei, Q.~Lyu, Q.~Chen, R.~Banjade, R.~E. Rudolph, R.~Gabriel, R.~Habacker, R.~Risco, R.~Milli{\`e}re, R.~Garg, R.~Barnes, R.~A. Saurous, R.~Arakawa, R.~Raymaekers, R.~Frank, R.~Sikand, R.~Novak, R.~Sitelew, R.~L. Bras, R.~Liu, R.~Jacobs, R.~Zhang, R.~Salakhutdinov, R.~A. Chi,
  S.~R. Lee, R.~Stovall, R.~Teehan, R.~Yang, S.~Singh, S.~M. Mohammad, S.~Anand, S.~Dillavou, S.~Shleifer, S.~Wiseman, S.~Gruetter, S.~R. Bowman, S.~S. Schoenholz, S.~Han, S.~Kwatra, S.~A. Rous, S.~Ghazarian, S.~Ghosh, S.~Casey, S.~Bischoff, S.~Gehrmann, S.~Schuster, S.~Sadeghi, S.~Hamdan, S.~Zhou, S.~Srivastava, S.~Shi, S.~Singh, S.~Asaadi, S.~S. Gu, S.~Pachchigar, S.~Toshniwal, S.~Upadhyay, S.~S. Debnath, S.~Shakeri, S.~Thormeyer, S.~Melzi, S.~Reddy, S.~P. Makini, S.-H. Lee, S.~Torene, S.~Hatwar, S.~Dehaene, S.~Divic, S.~Ermon, S.~Biderman, S.~Lin, S.~Prasad, S.~Piantadosi, S.~Shieber, S.~Misherghi, S.~Kiritchenko, S.~Mishra, T.~Linzen, T.~Schuster, T.~Li, T.~Yu, T.~Ali, T.~Hashimoto, T.-L. Wu, T.~Desbordes, T.~Rothschild, T.~Phan, T.~Wang, T.~Nkinyili, T.~Schick, T.~Kornev, T.~Tunduny, T.~Gerstenberg, T.~Chang, T.~Neeraj, T.~Khot, T.~Shultz, U.~Shaham, V.~Misra, V.~Demberg, V.~Nyamai, V.~Raunak, V.~V. Ramasesh, vinay~uday prabhu, V.~Padmakumar, V.~Srikumar, W.~Fedus, W.~Saunders, W.~Zhang, W.~Vossen,
  X.~Ren, X.~Tong, X.~Zhao, X.~Wu, X.~Shen, Y.~Yaghoobzadeh, Y.~Lakretz, Y.~Song, Y.~Bahri, Y.~Choi, Y.~Yang, Y.~Hao, Y.~Chen, Y.~Belinkov, Y.~Hou, Y.~Hou, Y.~Bai, Z.~Seid, Z.~Zhao, Z.~Wang, Z.~J. Wang, Z.~Wang, and Z.~Wu.
\newblock Beyond the imitation game: Quantifying and extrapolating the capabilities of language models.
\newblock \emph{Transactions on Machine Learning Research}, 2023.
\newblock ISSN 2835-8856.
\newblock URL \url{https://openreview.net/forum?id=uyTL5Bvosj}.

\bibitem[Sung et~al.(2022)Sung, Jeong, Choi, Kim, Lee, and Kang]{bern}
M.~Sung, M.~Jeong, Y.~Choi, D.~Kim, J.~Lee, and J.~Kang.
\newblock {BERN2:} an advanced neural biomedical named entity recognition and normalization tool.
\newblock \emph{Bioinformatics}, 38\penalty0 (20):\penalty0 4837–4839, Sept. 2022.
\newblock ISSN 1367-4811.
\newblock \doi{10.1093/bioinformatics/btac598}.
\newblock URL \url{http://dx.doi.org/10.1093/bioinformatics/btac598}.

\bibitem[Szklarczyk et~al.(2020)Szklarczyk, Gable, Nastou, Lyon, Kirsch, Pyysalo, Doncheva, Legeay, Fang, Bork, Jensen, and von Mering]{17}
D.~Szklarczyk, A.~L. Gable, K.~C. Nastou, D.~Lyon, R.~Kirsch, S.~Pyysalo, N.~T. Doncheva, M.~Legeay, T.~Fang, P.~Bork, L.~J. Jensen, and C.~von Mering.
\newblock {The STRING database in 2021: customizable protein–protein networks, and functional characterization of user-uploaded gene/measurement sets}.
\newblock \emph{Nucleic Acids Research}, 49\penalty0 (D1):\penalty0 D605--D612, 11 2020.
\newblock ISSN 0305-1048.
\newblock \doi{10.1093/nar/gkaa1074}.
\newblock URL \url{https://doi.org/10.1093/nar/gkaa1074}.

\bibitem[Touvron et~al.(2023)Touvron, Martin, Stone, Albert, Almahairi, Babaei, Bashlykov, Batra, Bhargava, Bhosale, Bikel, Blecher, Ferrer, Chen, Cucurull, Esiobu, Fernandes, Fu, Fu, Fuller, Gao, Goswami, Goyal, Hartshorn, Hosseini, Hou, Inan, Kardas, Kerkez, Khabsa, Kloumann, Korenev, Koura, Lachaux, Lavril, Lee, Liskovich, Lu, Mao, Martinet, Mihaylov, Mishra, Molybog, Nie, Poulton, Reizenstein, Rungta, Saladi, Schelten, Silva, Smith, Subramanian, Tan, Tang, Taylor, Williams, Kuan, Xu, Yan, Zarov, Zhang, Fan, Kambadur, Narang, Rodriguez, Stojnic, Edunov, and Scialom]{ll}
H.~Touvron, L.~Martin, K.~Stone, P.~Albert, A.~Almahairi, Y.~Babaei, N.~Bashlykov, S.~Batra, P.~Bhargava, S.~Bhosale, D.~Bikel, L.~Blecher, C.~C. Ferrer, M.~Chen, G.~Cucurull, D.~Esiobu, J.~Fernandes, J.~Fu, W.~Fu, B.~Fuller, C.~Gao, V.~Goswami, N.~Goyal, A.~Hartshorn, S.~Hosseini, R.~Hou, H.~Inan, M.~Kardas, V.~Kerkez, M.~Khabsa, I.~Kloumann, A.~Korenev, P.~S. Koura, M.-A. Lachaux, T.~Lavril, J.~Lee, D.~Liskovich, Y.~Lu, Y.~Mao, X.~Martinet, T.~Mihaylov, P.~Mishra, I.~Molybog, Y.~Nie, A.~Poulton, J.~Reizenstein, R.~Rungta, K.~Saladi, A.~Schelten, R.~Silva, E.~M. Smith, R.~Subramanian, X.~E. Tan, B.~Tang, R.~Taylor, A.~Williams, J.~X. Kuan, P.~Xu, Z.~Yan, I.~Zarov, Y.~Zhang, A.~Fan, M.~Kambadur, S.~Narang, A.~Rodriguez, R.~Stojnic, S.~Edunov, and T.~Scialom.
\newblock {Llama 2: Open Foundation and Fine-Tuned Chat Models}.
\newblock \emph{arXiv preprint arXiv:2307.09288}, 2023.

\bibitem[Wang et~al.(2019)Wang, Singh, Michael, Hill, Levy, and Bowman]{1}
A.~Wang, A.~Singh, J.~Michael, F.~Hill, O.~Levy, and S.~R. Bowman.
\newblock {GLUE}: A multi-task benchmark and analysis platform for natural language understanding.
\newblock In \emph{ICLR}, 2019.
\newblock URL \url{https://openreview.net/forum?id=rJ4km2R5t7}.

\bibitem[Wei et~al.(2022)Wei, Wang, Schuurmans, Bosma, brian ichter, Xia, Chi, Le, and Zhou]{wei2022chain}
J.~Wei, X.~Wang, D.~Schuurmans, M.~Bosma, brian ichter, F.~Xia, E.~H. Chi, Q.~V. Le, and D.~Zhou.
\newblock Chain of thought prompting elicits reasoning in large language models.
\newblock In \emph{NeurIPS}, 2022.
\newblock URL \url{https://openreview.net/forum?id=_VjQlMeSB_J}.

\bibitem[Xu et~al.(2022)Xu, Zhang, Lu, Zhu, Zhang, Chang, Liu, and Tang]{19}
M.~Xu, Z.~Zhang, J.~Lu, Z.~Zhu, Y.~Zhang, M.~Chang, R.~Liu, and J.~Tang.
\newblock {PEER: A Comprehensive and Multi-Task Benchmark for Protein Sequence Understanding}.
\newblock In \emph{NeurIPS}, volume~35, pages 35156--35173. Curran Associates, Inc., 2022.

\bibitem[Xu et~al.(2023)Xu, Yuan, Miret, and Tang]{xu_protst_2023}
M.~Xu, X.~Yuan, S.~Miret, and J.~Tang.
\newblock {P}rot{ST}: Multi-modality learning of protein sequences and biomedical texts.
\newblock In \emph{ICML}, volume 202, pages 38749--38767, 2023.
\newblock URL \url{https://proceedings.mlr.press/v202/xu23t.html}.

\bibitem[Xu et~al.(2021)Xu, Liu, Yi, Zhou, Li, and Wu]{8}
S.~Xu, Y.~Liu, X.~Yi, S.~Zhou, H.~Li, and Y.~Wu.
\newblock {Native Chinese Reader: A Dataset Towards Native-Level Chinese Machine Reading Comprehension}.
\newblock In \emph{NeurIPS Datasets and Benchmarks Track (Round 2)}, 2021.
\newblock URL \url{https://openreview.net/forum?id=GEcWUTN1v1v}.

\bibitem[Zhang et~al.(2024)Zhang, Li, Le, Shou, Xiong, and Sahoo]{moon}
D.~J. Zhang, D.~Li, H.~Le, M.~Z. Shou, C.~Xiong, and D.~Sahoo.
\newblock Moonshot: Towards controllable video generation and editing with multimodal conditions.
\newblock \emph{arXiv preprint arXiv:2401.01827}, 2024.

\bibitem[Zhou et~al.(2023)Zhou, Fu, Zhang, Cheng, and Yu]{zhou_protein_2023}
H.-Y. Zhou, Y.~Fu, Z.~Zhang, B.~Cheng, and Y.~Yu.
\newblock Protein representation learning via knowledge enhanced primary structure reasoning.
\newblock In \emph{ICLR}, 2023.
\newblock URL \url{https://openreview.net/forum?id=VbCMhg7MRmj}.

\end{thebibliography}
\newpage

\section*{Checklist}


\begin{enumerate}

\item For all authors...
\begin{enumerate}
  \item Do the main claims made in the abstract and introduction accurately reflect the paper's contributions and scope?
    \answerYes{Yes, they do.}
  \item Did you describe the limitations of your work?
    \answerYes{See Section~\ref{limitations}.}
  \item Did you discuss any potential negative societal impacts of your work?
    \answerNA{Our benchmark and dataset will not lead to any potential negative societal impacts}
  \item Have you read the ethics review guidelines and ensured that your paper conforms to them?
    \answerYes{We confirm that our paper conforms to them.}
\end{enumerate}

\item If you are including theoretical results...
\begin{enumerate}
  \item Did you state the full set of assumptions of all theoretical results?
    \answerNA{Our benchmark and dataset paper does not include any theoretical results.}
	\item Did you include complete proofs of all theoretical results?
    \answerNA{Our benchmark and dataset paper does not include any theoretical results.}
\end{enumerate}

\item If you ran experiments (e.g. for benchmarks)...
\begin{enumerate}
  \item Did you include the code, data, and instructions needed to reproduce the main experimental results (either in the supplemental material or as a URL)?
    \answerYes{}
  \item Did you specify all the training details (e.g., data splits, hyperparameters, how they were chosen)?
    \answerYes{See Appendix Section~\ref{training_details}.}
	\item Did you report error bars (e.g., with respect to the random seed after running experiments multiple times)?
    \answerNo{We did not conduct multiple experiments or set random seeds.}
	\item Did you include the total amount of compute and the type of resources used (e.g., type of GPUs, internal cluster, or cloud provider)?
    \answerYes{See Appendix Section~\ref{training_details}.}
\end{enumerate}

\item If you are using existing assets (e.g., code, data, models) or curating/releasing new assets...
\begin{enumerate}
  \item If your work uses existing assets, did you cite the creators?
    \answerYes{See Section~\ref{related_work}.}
  \item Did you mention the license of the assets?
    \answerYes{We provide the license in the dataset hugging face link.}
  \item Did you include any new assets either in the supplemental material or as a URL?
    \answerYes{See Section~\ref{related_work}.}
  \item Did you discuss whether and how consent was obtained from people whose data you're using/curating?
    \answerNA{Our dataset does not contain any personal information.}
  \item Did you discuss whether the data you are using/curating contains personally identifiable information or offensive content?
    \answerYes{See Appendix Section~\ref{ap_b2}}
\end{enumerate}

\item If you used crowdsourcing or conducted research with human subjects...
\begin{enumerate}
  \item Did you include the full text of instructions given to participants and screenshots, if applicable?
    \answerNA{We haven't used crowdsourcing or conducted research with human subjects.}
  \item Did you describe any potential participant risks, with links to Institutional Review Board (IRB) approvals, if applicable?
    \answerNA{We haven't used crowdsourcing or conducted research with human subjects.}
  \item Did you include the estimated hourly wage paid to participants and the total amount spent on participant compensation?
    \answerNA{We haven't used crowdsourcing or conducted research with human subjects.}
\end{enumerate}

\end{enumerate}

\newpage

\appendix
\section*{Appendix: A Finetuning Dataset and Benchmark for Large Language Models on Protein Sequence Understanding}

The questionnaire is based on the award paper of NeurIPS 2022 \cite{app}.
\section{Ethics statement}
This investigation adhered strictly to ethical guidelines and research best practices. All protein data utilized were sourced from publicly available datasets, and no proprietary or confidential information was employed. Stringent quality control measures and security checks have been implemented to eliminate any potential presence of harmful or malicious content in our dataset. It is important to acknowledge the profound implications and potential risks associated with integrating Large Language Model (LLM) and biomolecular knowledge. While our primary goal is to contribute positively to scientific understanding, we are cognizant of the potential misuse of these tools in the wrong hands. The combination of LLMs and protein sequence data could be exploited by malicious actors to generate harmful substances, such as illicit drugs. We strongly emphasize the imperative for users to adhere to the highest ethical standards when utilizing our dataset, promoting fairness, transparency, and responsibility in their research. Any utilization of the dataset that may result in harm or pose a detriment to society is strictly prohibited.

\section{Questionnaire}
\subsection{Motivation}
Q1 \textbf{For what purpose was the dataset created? Was there a specific task in mind?} Was there a specific gap that needed to be filled? Please provide a description.
\begin{enumerate}
\item[]
Our proposed \texttt{ProteinLMDataset} and \texttt{ProteinLMBench} address the limitations of existing resources, offering a comprehensive evaluation platform for natural language processing tasks involving Chinese-English text pairs, protein sequences, protein design, and analysis. The datasets can help large language models (LLMs), typically designed for natural language, effectively understand protein sequences as if they were a foreign language. By incorporating protein-related details and sequences in multiple languages, our dataset and benchmark aim to train LLMs to understand protein sequences in a cross-lingual manner.

\texttt{ProteinLMDataset} comprises 17.46 billion tokens for self-supervised pretraining, alongside a dataset of 893K instructions for supervised fine-tuning. Innovatively structured into three segments: 0.69\% Chinese-English text pairs in protein science, 41.51\% protein sequence-English text pairs, and 57.80\% protein-related English text, \texttt{ProteinLMDataset} can contribute to enhancing the cross-lingual capabilities of LLMs (English-Chinese-Sequence).

\texttt{ProteinLMBench} is the benchmark designed to evaluate the protein-related capabilities of LLMs. It has 944 multiple-choice questions, respectively, on protein tasks. \texttt{ProteinLMBench} is pioneering in its aim to thoroughly assess LLM performance in comprehending protein sequences.

\end{enumerate}
Q2 \textbf{Who created the dataset (e.g., which team, research group) and on behalf of which entity (e.g., company, institution, organization)?}
\begin{enumerate}
\item[] This dataset is jointly presented by Toursun Synbio with top research institutes Johns Hopkins University, University of Cambridge, Shanghai Institute for Biomedical and Pharmaceutical Technologies, Shanghai AI Laboratory, Shanghai Jiao Tong University, and UNSW Syndey. 

Toursun Synbio is an AI-driven protein design and manufacturing enterprise with its core team members coming from world-renowned universities. The company is committed to leading protein engineering with AI, shaping the future of synthetic biology, and addressing global health and environmental challenges.
\end{enumerate}
Q3 \textbf{Who funded the creation of the dataset? } If there is an associated grant, please provide the name of the grantor and the grant name and number.
\begin{enumerate}
\item[] This work was sponsored Toursun Synbio and Shanghai AI Laboratory.
\end{enumerate}
Q4 \textbf{Any other comments?}
\begin{enumerate}
\item[] No.
\end{enumerate}
\subsection{Composition} \label{ap_b2}
Q5 \textbf{What do the instances that comprise the dataset represent (e.g., documents, photos, people, countries)?} Are there multiple types of instances (e.g., movies, users, and ratings; people and interactions between them; nodes and edges)? Please provide a description.
\begin{enumerate}
\item[] \texttt{ProteinLMDataset} is innovatively structured into three segments: 0.69\% Chinese-English text pairs in protein science, 41.51\% protein sequence-English text pairs, and 57.80\% protein-related English text. \texttt{ProteinLMBench} has 944 multi-choice questions on protein tasks respectively.
\end{enumerate}
Q6 \textbf{How many instances are there in total (of each type, if appropriate)?}
\begin{enumerate}
\item[] \texttt{ProteinLMDataset} comprises 17.46 billion tokens for self-supervised pretraining, alongside a dataset of 893K instructions for supervised fine-tuning. \texttt{ProteinLMBench} has 944 multiple-choice questions respectively on protein-related tasks to evaluate model generalization. This large-scale dataset covers a wealth of protein knowledge and can support the development of large language models to advance natural language processing tasks related to proteins.
\end{enumerate}
Q7 \textbf{Does the dataset contain all possible instances or is it a sample (not necessarily
random) of instances from a larger set?} If the dataset is a sample, then what is the larger
set? Is the sample representative of the larger set (e.g., geographic coverage)? If so, please
describe how this representativeness was validated/verified. If it is not representative of the
larger set, please describe why not (e.g., to cover a more diverse range of instances, because
instances were withheld or unavailable).
\begin{enumerate}
\item[] The \texttt{ProteinLMDataset} is full dataset that we present. \texttt{ProteinLMBench} is randomly sampled from a dataset we created with over 100K protein related question and answer pairs. Although larger datasets may contain more possibilities, due to manpower limitations, it is difficult for us to scale up here.
\end{enumerate}
Q8 \textbf{What data does each instance consist of ? “Raw” data (e.g., unprocessed text or images)
or features?} In either case, please provide a description.

\begin{enumerate}
\item[] Data source for protein:

UniProt: \url{https://www.uniprot.org/help/downloads}

RefSeq: \url{https://ftp.ncbi.nlm.nih.gov/refseq/}

SIFTS: \url{https://ftp.ebi.ac.uk/pub/databases/msd/sifts/}

Enzyme Nomenclature: \url{https://iubmb.qmul.ac.uk/enzyme/}

Data sources for academic papers and journals:

PubMed Central: \url{https://www.ncbi.nlm.nih.gov/pmc/tools/ftp/}

PubMed: \url{https://pubmed.ncbi.nlm.nih.gov/download/}
\end{enumerate}

Q9 \textbf{Is there a label or target associated with each instance?} If so, please provide a description.
\begin{enumerate}
\item[] For SSL dataset there is no label. For SFT dataset each data has the answer to the question as the label.
\end{enumerate}
Q10 \textbf{Is any information missing from individual instances?} If so, please provide a description, explaining why this information is missing (e.g., because it was unavailable). This does
not include intentionally removed information, but might include, e.g., redacted text.
\begin{enumerate}
\item[] No.
\end{enumerate}
Q11 \textbf{Are relationships between individual instances made explicit (e.g., users’ movie
ratings, social network links)?} If so, please describe how these relationships are made
explicit.
\begin{enumerate}
\item[] No.
\end{enumerate}
Q12 \textbf{Are there recommended data splits (e.g., training, development/validation, test-
ing)?} If so, please provide a description of these splits, explaining the rationale behind them.
\begin{enumerate}
\item[] No.
\end{enumerate}
Q13 \textbf{Are there any errors, sources of noise, or redundancies in the dataset?} If so, please
provide a description.
\begin{enumerate}
 \item[] There exist near duplicate protein sequences which makes possible a good many to one embedding in certain scenarios. When attaching literature to a protein sequence, two sequences may refer to the same paper that mentioned both sequences.
\end{enumerate}
Q14 \textbf{Is the dataset self-contained, or does it link to or otherwise rely on external
resources (e.g., websites, tweets, other datasets)?} If it links to or relies on external
resources, a) are there guarantees that they will exist, and remain constant, over time; b) are
there official archival versions of the complete dataset (i.e., including the external resources as
they existed at the time the dataset was created); c) are there any restrictions (e.g., licenses,
fees) associated with any of the external resources that might apply to a future user? Please
provide descriptions of all external resources and any restrictions associated with them, as well
as links or other access points, as appropriate.
\begin{enumerate}
\item[] The dataset is self-contained as the protein sequences and literature are downloaded from the academic datasets such as UniProt, NCBI and PubMed, and then cleaned and well preprocessed.
\end{enumerate}
Q15 \textbf{Does the dataset contain data that might be considered confidential (e.g., data
that is protected by legal privilege or by doctor–patient confidentiality, data that
includes the content of individuals’ non-public communications)?} If so, please provide
a description.
\begin{enumerate}
\item[] This dataset was collected using openly available parts of the internet with the assumption
that any data found was intended to be shared freely. 
\end{enumerate}
Q16 \textbf{Does the dataset contain data that, if viewed directly, might be offensive, insulting,
threatening, or might otherwise cause anxiety?} If so, please describe why.
\begin{enumerate}
\item[] The dataset is the biological information and references of proteins and there is no such concern.
\end{enumerate}
Q17 \textbf{Does the dataset relate to people?} If not, you may skip the remaining questions in this
section.
\begin{enumerate}
\item[] No.
\end{enumerate}
Q18 \textbf{Does the dataset identify any subpopulations (e.g., by age, gender)?}
\begin{enumerate}
\item[] No
\end{enumerate}
Q19 \textbf{Is it possible to identify individuals (i.e., one or more natural persons), either
directly or indirectly (i.e., in combination with other data) from the dataset?} If
so, please describe how.
\begin{enumerate}
\item[] No.
\end{enumerate}
Q20 \textbf{Does the dataset contain data that might be considered sensitive in any way
(e.g., data that reveals racial or ethnic origins, sexual orientations, religious be-
liefs, political opinions or union memberships, or locations;} financial or health
data; biometric or genetic data; forms of government identification, such as social
security numbers; criminal history)? If so, please provide a description.
\begin{enumerate}
\item[] The protein information contains biological categories which the protein belongs to. But it is not related to racial or ethnic origins, sexual orientions and so on.
\end{enumerate}
Q21 \textbf{Any other comments?}
\begin{enumerate}
\item[] No
\end{enumerate}
\subsection{Collection Process}
Q22 \textbf{ How was the data associated with each instance acquired?} Was the data directly
observable (e.g., raw text, movie ratings), reported by subjects (e.g., survey responses), or
indirectly inferred/derived from other data (e.g., part-of-speech tags, model-based guesses for
age or language)? If data was reported by subjects or indirectly inferred/derived from other
data, was the data validated/verified? If so, please describe how.

\begin{enumerate}
\item[] None of the journal text and protein sequence parts in the \texttt{ProteinLMDataset} SSL part have been processed. However, during the concatenation of these two parts, NER and NEN techniques were applied, with detailed procedures described in the main text. For the UniProtKB Swiss-Prot part, the downloaded data was filled according to a predefined template.In \texttt{ProteinLMDataset} SFT part, the data was generated by creating suitable templates with a Large Language Model (LLM) and then filling in the corresponding data. 

The \texttt{ProteinLMBench} dataset, used for evaluating the effectiveness of the large model, was generated by using the LLM to read the downloaded research papers and create relevant questions. These questions were then validated by both machines and humans. The specific process is detailed in the main text.
\end{enumerate}

Q23 \textbf{What mechanisms or procedures were used to collect the data (e.g., hardware
apparatus or sensor, manual human curation, software program, software API)?}
How were these mechanisms or procedures validated?
\begin{enumerate}
\item[] We ran a preprocessing script in python, over dozens of CPU cores, and 8 Nvidia 4090
cards. They were validated by python program we wrote.
\end{enumerate}
Q24 \textbf{If the dataset is a sample from a larger set, what was the sampling strategy (e.g.,
deterministic, probabilistic with specific sampling probabilities)?}
\begin{enumerate}
\item[] We deleted some literature that is not related to the protein.
\end{enumerate}
Q25 \textbf{Who was involved in the data collection process (e.g., students, crowdworkers,
contractors) and how were they compensated (e.g., how much were crowdworkers
paid)?}
\begin{enumerate}
\item[] No crowdworkers were used in the curation of the dataset. Only authors of the paper enabled its creation.
\end{enumerate}
Q26 \textbf{Over what timeframe was the data collected? Does this timeframe match the
creation timeframe of the data associated with the instances (e.g., recent crawl of
old news articles)?} If not, please describe the timeframe in which the data associated with
the instances was created.
\begin{enumerate}
\item[] The data collected from the database has the same timeframe as when the database is created. The literature may date back to 100 years ago.
\end{enumerate}
Q27 \textbf{Were any ethical review processes conducted (e.g., by an institutional review
board)?} If so, please provide a description of these review processes, including the outcomes,
as well as a link or other access point to any supporting documentation.
\begin{enumerate}
\item[] We corresponded with the University of Washington’s Human Subject Division, and as
we do not intervene with the people depicted in the data as well as the data being public,
they stated that the work did not require IRB review. Furthermore, the NeurIPS ethics
review determined that the work has no ethical issues.
\end{enumerate}
Q28 \textbf{Does the dataset relate to people?} If not, you may skip the remaining questions in this
section.
\begin{enumerate}
\item[] The protein sequence may be used by human being as protein the basic unit of life.
\end{enumerate}
Q29 \textbf{Did you collect the data from the individuals in question directly, or obtain it via
third parties or other sources (e.g., websites)?}
\begin{enumerate}
\item[]We retrieve the data from public databases including PubMed, UniProt and NCBI, and from the official website of Chinese journals.
\end{enumerate}
Q30 \textbf{Were the individuals in question notified about the data collection?} If so, please
describe (or show with screenshots or other information) how notice was provided, and provide
a link or other access point to, or otherwise reproduce, the exact language of the notification
itself.
\begin{enumerate}
\item[] Individuals were not notified about the data collection.
\end{enumerate}
Q31 \textbf{Did the individuals in question consent to the collection and use of their data?} If
so, please describe (or show with screenshots or other information) how consent was requested
and provided, and provide a link or other access point to, or otherwise reproduce, the exact
language to which the individuals consented.
\begin{enumerate}
\item[] The database and journals are public and free to download and use.
\end{enumerate}
Q32 \textbf{If consent was obtained, were the consenting individuals provided with a mechanism to revoke their consent in the future or for certain uses?} If so, please provide
a description, as well as a link or other access point to the mechanism (if appropriate).
\begin{enumerate}
\item[] Not applicable.
\end{enumerate}
Q33 \textbf{Has an analysis of the potential impact of the dataset and its use on data subjects
(e.g., a data protection impact analysis) been conducted?} If so, please provide a
description of this analysis, including the outcomes, as well as a link or other access point to
any supporting documentation.
\begin{enumerate}
\item[] The dataset we processed is used to assign and evaluate the capability of understanding proteins for foundation models. No ethical issue will be involved.
\end{enumerate}
Q34 \textbf{Any other comments?}
\begin{enumerate}
\item[] No
\end{enumerate}
\subsection{Preprocessing, Cleaning, and/or Labeling}
Q35 \textbf{Was any preprocessing/cleaning/labeling of the data done (e.g., discretization or
bucketing, tokenization, part-of-speech tagging, SIFT feature extraction, remova of instances, processing of missing values)?} If so, please provide a description. If not,
you may skip the remainder of the questions in this section.
\begin{enumerate}
\item[] Yes. Preprocessing/cleaning/labeling is conducted when the raw data is obtained from publica database for protein (PubMed, UniProt, NCBI and Chinese journals). Preprocessing/cleaning/labeling is done without any bias.
\end{enumerate}
Q36 \textbf{Was the “raw” data saved in addition to the preprocessed/cleaned/labeled data
(e.g., to support unanticipated future uses)?} If so, please provide a link or other access
point to the “raw” data.
\begin{enumerate}
\item[] We save the raw data in our local server. The data are not sensitive and will be repeated used in our study.
\end{enumerate}
Q37 \textbf{Is the software used to preprocess/clean/label the instances available?} If so, please
provide a link or other access point.

\begin{enumerate}
\item[] None of the preprocessing/cleaning/labeling was done using external software; the entire process was completed using custom Python scripts.
\end{enumerate}

Q38 \textbf{Any other comments?}
\begin{enumerate}
\item[] No.
\end{enumerate}
\subsection{Uses}
Q39 \textbf{Has the dataset been used for any tasks already?} If so, please provide a description.
\begin{enumerate}
\item[] It has been used for research related to proteins. In particular, we use \texttt{ProteinLMDataset} for training a protein language model and \texttt{ProteinLMBench} for evaluating a variety of LLMs in protein tasks.
\end{enumerate}
Q40 \textbf{Is there a repository that links to any or all papers or systems that use the dataset?}
If so, please provide a link or other access point.
\begin{enumerate}
\item[] Yes, the datasets can be downloaded from \url{https://huggingface.co/datasets/tsynbio/ProteinLMBench}.
\end{enumerate}
Q41 \textbf{What (other) tasks could the dataset be used for?}
\begin{enumerate}
\item[] Our dataset \texttt{ProteinLMDataset} can be used for training an LLM to linking the protein language and natural language (including English and Chinese). The dataset \texttt{ProteinLMBench} is used for evaluating the capability of LLMs in protein tasks including Protein general knowledge, 
Protein literature understanding,
Protein property prediction,
Protein function prediction,
Protein design.
\end{enumerate}
Q42 \textbf{Is there anything about the composition of the dataset or the way it was collected and preprocessed/cleaned/labeled that might impact future uses?} For example,
is there anything that a future user might need to know to avoid uses that could result in
unfair treatment of individuals or groups (e.g., stereotyping, quality of service issues) or other
undesirable harms (e.g., financial harms, legal risks) If so, please provide a description. Is
there anything a future user could do to mitigate these undesirable harms?
\begin{enumerate}
\item[] The protein sequence and information that we collected from the database may be updated, and new literature about a particular protein may appear in future. Future research can update the dataset with the same preprocessing/cleaning/labeling method that we used while the method per se has no bias toward any individual or species that may violate ethical rule.
\end{enumerate}
Q43 \textbf{Are there tasks for which the dataset should not be used?} If so, please provide a
description.
\begin{enumerate}
\item[] No. The dataset is for protein related tasks, specified for LLMs supporting Chinese and English. It can be also used for LLMs supporting multilanguage. 
\end{enumerate}
Q44 \textbf{Any other comments?}
\begin{enumerate}
\item[] No.
\end{enumerate}
\subsection{Distribution}
Q45 \textbf{Will the dataset be distributed to third parties outside of the entity (e.g., company, institution, organization) on behalf of which the dataset was created? }If so, please
provide a description.
\begin{enumerate}
\item[] Yes, the dataset is open-source.
\end{enumerate}
Q46 \textbf{How will the dataset be distributed (e.g., tarball on website, API, GitHub)? Does
the dataset have a digital object identifier (DOI)?}
\begin{enumerate}
\item[] The data is available through Huggingface datasets, and can be downloaded from \url{https://huggingface.co/datasets/tsynbio/ProteinLMBench}.
\end{enumerate}
Q47 \textbf{When will the dataset be distributed?}
\begin{enumerate}
\item[] 03/06/2024 and onward.
\end{enumerate}
Q48 \textbf{Will the dataset be distributed under a copyright or other intellectual property
(IP) license, and/or under applicable terms of use (ToU)?} If so, please describe this
license and/or ToU, and provide a link or other access point to, or otherwise reproduce, any
relevant licensing terms or ToU, as well as any fees associated with these restrictions.

\begin{enumerate}
\item[] Apache License 2.0
\end{enumerate}

Q49 \textbf{Have any third parties imposed IP-based or other restrictions on the data associated with the instances?} If so, please describe these restrictions, and provide a link or
other access point to, or otherwise reproduce, any relevant licensing terms, as well as any fees
associated with these restrictions.
\begin{enumerate}
\item[] Toursun Synbio owns the metadata and release as CC-BY-4.0.
\item[] We do not own the copyright of the protein sequence or the related literature.
\end{enumerate}
Q50 \textbf{Do any export controls or other regulatory restrictions apply to the dataset or to
individual instances?} If so, please describe these restrictions, and provide a link or other
access point to, or otherwise reproduce, any supporting documentation.
\begin{enumerate}
\item[] No.
\end{enumerate}
Q51 \textbf{Any other comments?}
\begin{enumerate}
\item[] No.
\end{enumerate}
\subsection{Maintenance}
Q52 \textbf{Who will be supporting/hosting/maintaining the dataset?}
\begin{enumerate}
\item[] Huggingface will support hosting of the metadata.
\item[] The Toursun Synbio supports hosting of the embeddings and backups of the rest, and will maintain the samples distributed.
\end{enumerate}
Q53 \textbf{How can the owner/curator/manager of the dataset be contacted (e.g., email
address)?}
\begin{enumerate}
\item[] If any questions, please email to \texttt{contact@tsynbio.com}.
\end{enumerate}
Q54 \textbf{Is there an erratum?} If so, please provide a link or other access point.
\begin{enumerate}
\item[] There is no erratum for our initial release. Errata will be documented as future releases
on the dataset website.
\end{enumerate}
Q55 \textbf{Will the dataset be updated (e.g., to correct labeling errors, add new instances,
delete instances)?} If so, please describe how often, by whom, and how updates will be
communicated to users (e.g., mailing list, GitHub)?
\begin{enumerate}
\item[] \texttt{ProteinLMDataset} will not be updated. However a future Toursun Synbio-streamed-from-CC may exist
for updates. \texttt{ProteinLMBench} will be updated with more multi-choice questions on protein tasks.
\end{enumerate}
Q56 \textbf{If the dataset relates to people, are there applicable limits on the retention of the
data associated with the instances (e.g., were individuals in question told that
their data would be retained for a fixed period of time and then deleted)?} If so,
please describe these limits and explain how they will be enforced.
\begin{enumerate}
\item[] The datasets are not related to people. But people may contact us at the \texttt{contact@tsynbio.com}, and  add specific samples to a blacklist.
\end{enumerate}
Q57 \textbf{Will older versions of the dataset continue to be supported/hosted/maintained?}
If so, please describe how. If not, please describe how its obsolescence will be communicated to
users.
\begin{enumerate}
\item[] This the first version of the datasets.
\end{enumerate}
Q58 \textbf{If others want to extend/augment/build on/contribute to the dataset, is there a
mechanism for them to do so?} If so, please provide a description. Will these contribu-
tions be validated/verified? If so, please describe how. If not, why not? Is there a process
for communicating/distributing these contributions to other users? If so, please provide a
description.
\begin{enumerate}
\item[] No.
\end{enumerate}
Q59 \textbf{Any other comments?}
\begin{enumerate}
\item[] No.
\end{enumerate}


\section{Experiments and training details} \label{training_details}
\subsection{LLMs Instruction}
TourSynbio is a model developed by building upon the InternLM2 through further pre-training and supervised fine-tuning. It retains the original architectural design without any modifications.
\subsection{Self-supervised training \& Supervised fine-tune training details}
\subsubsection{Hyperparameter Configuration}
Common hyperparameters for both SSL and SFT phases included:

\textbf{Learning Rate}: 0.0002 

\textbf{Batch Size}: 16

\textbf{Optimizer}: AdamW

\textbf{Parameter Scheduler}: Alternating between LinearLR and CosineAnnealingLR

\textbf{Training Method}: LoRA (LoRA Alpha = 16, LoRA Dropout = 0.1)

\subsubsection{Self-Supervised Learning (SSL) Phase}
For the SSL phase, we initialized our model weights based on the InternLM2 model. We trained on a mixed dataset comprising 50 billion tokens, integrating both our specific data and a broad range of general-purpose data. The training utilized a batch size of 16 and a fixed learning rate of 0.0002. The same process followed for the other LLM models is highlighted in the main manuscript.
\subsubsection{Supervised Fine-Tuning (SFT) Phase}
Post-pre-training, the model underwent SFT using 850,000 samples from our constructed dataset. The fine-tuning process spanned three additional epochs, refining the weights to achieve optimal performance.

\subsection{Model performance comparison \& Further analysis}

Tab.~\ref{tab:comparison} presents the analytical accuracy scores of the different LLM models for the self-supervise training (SSL) and the supervised fine-tuning training (SFT) experiments.

\begin{table}
  \caption{Different protein language models accuracy score and inference time in the evaluation cohort.}
  \label{tab:comparison}
  \centering
  \begin{tabular}{lcc}
    \toprule
    \multicolumn{3}{c}{Protein models score}                   \\
    \cmidrule(r){1-3}
    Model & Correct Rate (\%) & Inference Time (minutes)\\
    \midrule
    GPT4.0-turbo                & 57.94 & 15.52\\
    InternLM2-20B               & 57.52 & 47.20\\    
    GPT3.5-turbo                & 55.19 & 21.03\\
    InternLM2-7B                & 54.98 & 19.23\\
    InternLM2-Chat-7B           & 54.76 & 35.58\\
    InternLM2-Chat-20B          & 51.38 & 31.11\\
    Yi-6B                       & 50.85 & 59.05\\
    Mistral-7B-Instruct-v0.2    & 50.11 & 13.00\\
    ChatGLM3-6B                 & 48.94 & 8.00\\
    Baichuan2-7B                & 44.49 & 16.37\\
    InternLM-Chat-20B           & 40.54 & 66.00\\   
    Llama2-7B                   & 39.64 & 64.00\\ 
    Moonshot                    & 38.26 & 16.25\\
    Qwen1.5-7B                  & 21.73 & 13.00\\
    Falcon-7B-Instruct          & 20.55 & 25.42\\
    Falcon-7B                   & 19.17 & 15.55\\ \midrule[0.5pt]
    InternLM2-Protein-7B        & \pmb{62.18} &22.34\\
    InternLM2-Protein-7B (w/o SSL) & 58.26 & 21.36\\

    \bottomrule
  \end{tabular}
\end{table}

\end{document}